%% file: mmSpeech_IMWUT.tex
\documentclass[acmlarge]{acmart}
\AtBeginDocument{%
  }

\setcopyright{acmlicensed}
\copyrightyear{2025}
\acmYear{2025}
\acmDOI{10.1145/3770708}

\acmJournal{IMWUT}
\acmVolume{9}
\acmNumber{4}
\acmArticle{179}
\acmMonth{12}

\usepackage{xspace}
\usepackage{subfigure}
\usepackage{graphicx}
\usepackage{balance}
\usepackage{comment}
\usepackage{alltt}
\usepackage{multirow}
\usepackage{CJK}
\usepackage{textcomp}
\usepackage{colortbl}
\usepackage{amsmath}
\usepackage{amsfonts}
\usepackage{url}
\usepackage{algorithmic}
\usepackage[ruled,vlined]{algorithm2e}
\usepackage{multicol}
\usepackage{lipsum}
\usepackage{mathrsfs}
\usepackage{overpic}
\usepackage{booktabs}
\usepackage{bbm}
\usepackage{makecell}
\usepackage{hyperref}
\usepackage{hyperxmp}

\newcommand{\ie}{\emph{i.e.}\xspace}
\newcommand{\eg}{\emph{e.g.}\xspace}

\newcommand{\etal}{\emph{et al.}\xspace}

\newcommand{\Figure}[1]{Fig.~\ref{fig:#1}}
\newcommand{\Equation}[1]{Eq.~\ref{eq:#1}}
\newcommand{\Section}[1]{\S\ref{sec:#1}}
\newcommand{\Table}[1]{Table~\ref{tab:#1}}

\usepackage{booktabs}

\newcommand\system{mmSpeech\xspace}

\begin{document}
\title{We Can Hear You with mmWave Radar! An End-to-End Eavesdropping System}

\author{Dachao Han}
\authornote{Both authors contributed equally to this research.}
\email{handachao@stu.xjtu.edu.cn}
\orcid{0009-0007-2414-338X}
\author{Teng Huang}
\authornotemark[1]
\orcid{0009-0000-0910-5739}
\email{huangteng990920@stu.xjtu.edu.cn}
\affiliation{%
  \institution{Xi'an Jiaotong University}
    \city{Xi'an}
  \country{China}
}

% \author{Dachao Han*}
% \orcid{0009-0007-2414-338X}
% \email{handachao@stu.xjtu.edu.cn}
% \affiliation{%
%   \institution{Xi'an Jiaotong University}
%     \city{Xi'an}
%   \country{China}
%   }

% \author{Teng Huang*}
% \orcid{0009-0000-0910-5739}
% \email{huangteng990920@stu.xjtu.edu.cn}
% \affiliation{%
%   \institution{Xi'an Jiaotong University}
%   \country{China}
%   \authornote{These authors contributed equally to this work.}
% }

\author{Han Ding}
\authornotemark[2]
\orcid{0000-0002-5274-7988}
\email{dinghan@xjtu.edu.cn}
\affiliation{%
  \institution{Xi'an Jiaotong University}
  \country{China}
  \authornote{Han Ding is the Corresponding author.}
}

\author{Cui Zhao}
\orcid{0000-0002-4603-4914}
\email{zhaocui@xjtu.edu.cn}
\affiliation{%
  \institution{Xi'an Jiaotong University}
  \country{China}}

\author{Fei Wang}
\orcid{0000-0002-0750-6990}
\email{feymanw@xjtu.edu.cn}
\affiliation{%
  \institution{Xi'an Jiaotong University}
  \country{China}}

\author{Ge Wang}
\orcid{0000-0002-3845-1646}
\email{gewang@xjtu.edu.cn}
\affiliation{%
  \institution{Xi'an Jiaotong University}
  \country{China}}

\author{Wei Xi}
\orcid{0000-0001-9348-2982}
\email{xiwei@xjtu.edu.cn}
\affiliation{%
  \institution{Xi'an Jiaotong University}
  \country{China}}

\renewcommand{\shortauthors}{Han et al.}

% \linespread{0.99}

\begin{abstract}
With the rise of voice-enabled technologies, loudspeaker playback has become widespread, posing increasing risks to speech privacy. Traditional eavesdropping methods often require invasive access or line-of-sight, limiting their practicality. In this paper, we present \system, an end-to-end mmWave-based eavesdropping system that reconstructs intelligible speech solely from vibration signals induced by loudspeaker playback, even through walls and without prior knowledge of the speaker.
To achieve this, we reveal an optimal combination of vibrating material and radar sampling rate for capturing high-quality vibrations using narrowband mmWave signals. We then design a deep neural network that reconstructs intelligible speech from the estimated noisy spectrograms. To further support downstream speech understanding, we introduce a synthetic training pipeline and selectively fine-tune the encoder of a pre-trained ASR model.
We implement \system with a commercial mmWave radar and validate its performance through extensive experiments. Results show that \system achieves state-of-the-art speech quality and generalizes well across unseen speakers and various conditions.
\end{abstract}




\maketitle

\input{Body/intro}

\input{Body/related}

\input{Body/background}

\input{Body/overview}

\input{Body/sensing}

\input{Body/model}

\input{Body/implementation}

\input{Body/evaluation}

\input{Body/discussion}

\input{Body/cons}

\subsection*{Acknowledgements}
This work was supported by the NSFC Grant No. 62372365, 62302383, 62472346, and project funded by China Postdoctoral Science Foundation No.2023M742792.

% \newpage
% \IEEEtriggeratref{16}
\bibliographystyle{ACM-Reference-Format}
\bibliography{reference}

\appendix
\setcounter{equation}{0}
\renewcommand{\theequation}{A\arabic{equation}}
\input{Body/appendix.tex}

\end{document}

%% file: Body/intro.tex
\section{Introduction}

With the widespread adoption of remote conferencing systems, smart speakers, and voice-enabled devices, spoken audio is increasingly broadcast through loudspeakers in meeting rooms, offices, and homes. While these technologies enhance communication and convenience, they also introduce significant security and privacy risks. In particular, eavesdropping on loudspeaker playback presents a serious threat \cite{wei2015acoustic}, as it can lead to the unauthorized disclosure of sensitive content such as business strategies, confidential negotiations, and private conversations—without the speaker's awareness.

Traditional eavesdropping methods—including directional microphones \cite{Directional_Microphones}, spyware \cite{spyware}, or compromised IoT devices \cite{iotDevice}, often require line-of-sight (LOS) access or invasive techniques like device hijacking. These methods are typically complex, intrusive, and prone to detection. In response, researchers have explored more covert alternatives that capture speech via sound-induced vibrations in surrounding objects.
Several sensing modalities have been investigated for vibration-based speech recovery. Optical methods \cite{walker2022laser}\cite{nassi2022lamphone}, such as high-speed video and laser vibrometry, can detect minute surface motions but are constrained by LOS requirements and lighting conditions. 
Inertial sensors like accelerometers \cite{ba2020learning}\cite{hu2022accear} require direct contact with the vibrating object, restricting their practicality in real-world scenarios.
In contrast, wireless sensing—particularly using mmWave radar—offers a compelling solution. The mmWave radar can detect micro-vibrations from a distance, penetrate non-metallic obstructions, and operate passively without direct access to the audio source. 
Recent advances have demonstrated its precision in capturing fine-grained motion for applications such as vital sign monitoring and human activity recognition \cite{wang2022your}\cite{wang2024xrf55}\cite{xue2023towards}\cite{yuance2023mmyodar}\cite{ding2023mi}\cite{huang2025one}. Building on these capabilities, researchers have started exploring mmWave radar for vibration-based speech recovery. However, existing efforts \cite{basak2022mmspy}\cite{wang2022mmeve}\cite{wang2024vibspeech} remain limited: some can only reconstruct simple speech elements (\eg, digits or letters), others fail to operate through walls, or rely on pre-recorded voice samples of the victim, all of which reduce their effectiveness in real-world, stealthy eavesdropping.

To address these limitations, we propose \system, an end-to-end mmWave-based eavesdropping system capable of reconstructing intelligible speech solely from vibration signals induced by loudspeaker playback. Unlike prior methods, \system is designed to work without prior knowledge of the speaker's voice and can recover unconstrained speeches, making it a practical and effective solution for covert surveillance. To achieve this, we address two key technical challenges:

1) \textit{Fine-Grained Vibration Sensing}. The quality of vibration sensing heavily depends on the choice of vibrating material. We conduct a systematic analysis of candidate surfaces and identify a passive film with both a high bandwidth-to-mass ratio and strong mmWave reflectivity. These properties make it ideal for capturing subtle vibrations. Additionally, we challenge the common assumption that higher radar sampling rates yield better results. Considering radar chirp configurations, we determine and adopt a suitable sampling rate that maximizes the sensing resolution. 
Taking into account the narrowband nature of mmWave-based vibration sensing \cite{wang2024vibspeech}\cite{li2022mmphone} and these two key design choices, we demonstrate that capturing vibrations below 4 kHz can support reliable speech eavesdropping in our experimental setup.

2) \textit{End-to-End Speech Reconstruction}. To transform the noisy mmWave-derived vibration signals into intelligible speech, we design an end-to-end deep neural network (DNN) with a generator-discriminator architecture. The generator refines distorted vibration spectrograms and complements wideband components, while the discriminator promotes speech intelligibility. To avoid dependence on speaker-specific training data, we also propose a signal synthesis method that simulates mmWave vibration patterns using open-source speech and mixed noise models to further enhance generalization.

\begin{figure}[t]
\centering
\begin{minipage}[t]{\linewidth}
\centering
{\includegraphics[width=.85\textwidth]{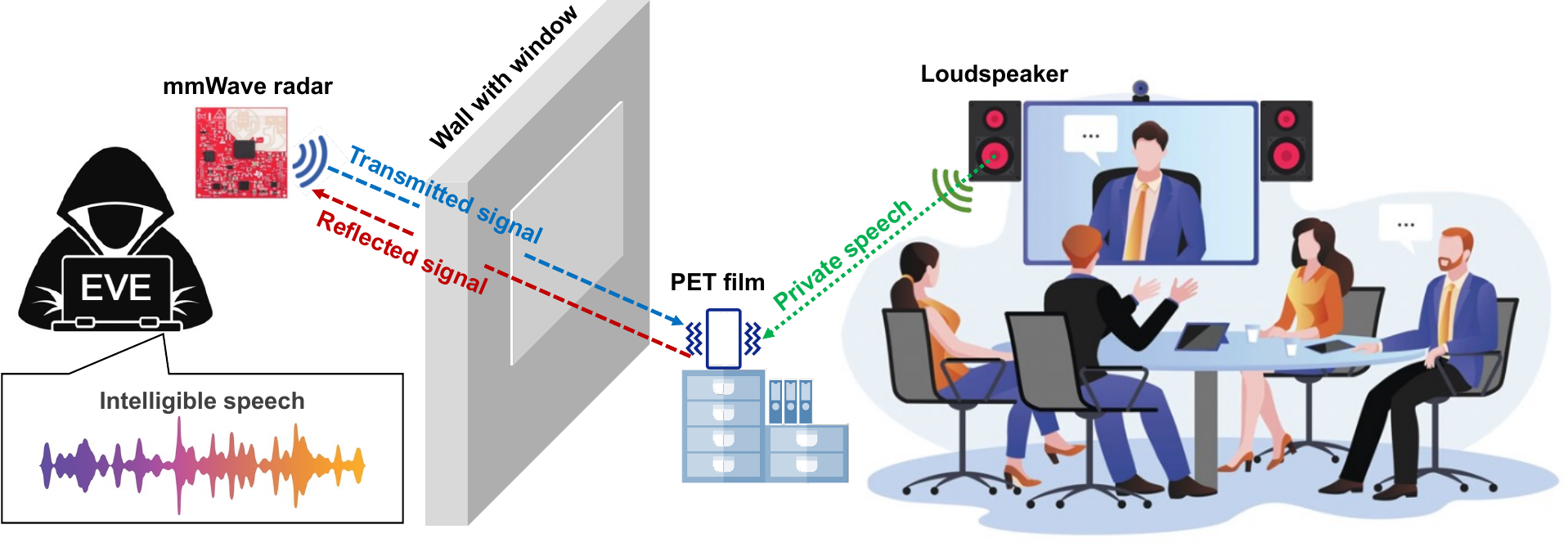}}
\caption{An eavesdropper equipped with a portable mmWave radar can covertly reconstruct private speech occurring inside a soundproof room. By capturing subtle vibration signals induced by loudspeaker playback, the attacker can not only recover intelligible speech but also transcribe its content using an ASR system—posing a serious threat to speech privacy even in acoustically isolated environments.}
\label{fig:intro}
\end{minipage}
\vspace{-0.15in}
\end{figure}

To evaluate \system, we build a prototype using a commercial off-the-shelf mmWave radar (TI IWR6843ISK) and construct a dataset comprising 2,400 speech clips, randomly sampled from the LibriSpeech corpus \cite{panayotov2015librispeech} and spoken by 47 distinct speakers. The clips are played through a commercial JBL loudspeaker to simulate playback conditions in practice. Through extensive experiments, we demonstrate that \system outperforms state-of-the-art (SOTA) methods. Our main contributions are summarized as follows:

$\bullet$ We reveal how material properties and radar sampling rates affect vibration sensing quality, enabling more effective capture of vibrations from narrowband mmWave signals.

$\bullet$ We design a novel end-to-end DNN that can reconstruct high-quality speech solely from mmWave vibration data, even in through-wall scenarios.

$\bullet$ We conduct extensive experiments showing that \system achieves SOTA performance across standard speech quality metrics (FWSegSNR: 9.43 dB, STOI: 0.80, MCD: 5.18, MEL: 2.09) and generalizes well across unseen speech contents, unseen speakers, and various environmental conditions.

%% file: Body/related.tex
\section{Related Work} \label{sec:related}
This paper investigates speech reconstruction and eavesdropping through vibration signals. Such speech-related vibrations can be captured using various sensing modalities, including optical sensors, inertial measurement units (IMUs), and wireless signals.

\subsection{Sensor-based Methods}
A number of studies focused on optical and IMU sensing to extract vibration cues related to audio. For instance, Davis \etal \cite{davis2014visual} demonstrated that a high-speed camera could detect sound-induced sub-pixel vibrations in everyday objects, enabling audio waveform reconstruction from videos.
Similarly, Nassi \etal proposed Lamphone \cite{nassi2022lamphone}, a system that monitored the minute oscillations of a desk lamp's bulb using a photodetector, achieving remote speech reconstruction from distances of up to 35 meters.
On the other hand, researchers also explored IMUs embedded in consumer devices to detect subtle motion caused by nearby sounds. AccelEve \cite{ba2020learning} analyzed vibrations from a smartphone's own speaker via its internal accelerometer to recover spoken words and infer speaker characteristics.
Building on this, AccEar \cite{hu2022accear} employed generative adversarial networks (GAN) to synthesize intelligible speech from low-sample-rate IMU data. Despite these methods are effective, sensor-based techniques are limited: optical sensors cannot operate through opaque barriers, while IMU-based methods require physical contact with the sound source, constraining their applicability in real-world, non-intrusive scenarios.

\subsection{Wireless-based Methods}
Wireless technologies have recently gained attention for speech recovery due to their ability to detect micro vibrations and penetrate physical obstructions. Among these, mmWave systems \cite{hu2023mmecho,basak2022mmspy,wang2022mmeve,feng2023mmeavesdropper,wang2024vibspeech,wang2022wavesdropper,lin2023high,xu2024mmear} are appealing due to their high spatial resolution enabled by shorter wavelengths. mmEcho \cite{hu2023mmecho} introduced a phase alignment method to accurately extract sound-induced vibrations from mmWave signals, enabling reliable speech reconstruction across different spatial positions and orientations.
Later works such as mmSpy \cite{basak2022mmspy} and mmEve \cite{wang2022mmeve} focused on decoding speech during phone calls by capturing vibrations from the smartphone's earpiece using mmWave radar. mmEavesdropper \cite{feng2023mmeavesdropper} further explored directional eavesdropping by introducing a signal augmentation strategy tailored to mmWave input.
Although these methods are effective, they are generally limited to recognizing simple speech elements like digits or letters and often struggle to operate effectively through walls. To overcome these limitations, the SOTA approach, VibSpeech \cite{wang2024vibspeech}, fused short speech segments with mmWave-captured vibrations to reconstruct high-quality wideband speech (up to 8 KHz). However, this approach depends on having a prior recording of the victim speaker's voice—requiring the system to pre-identify and collect speech data from the intended subject. This constraint significantly hinders its applicability in real-world, generalized eavesdropping scenarios.
In contrast, our proposed system removes the need for any pre-recorded voice data. It reconstructs speech exclusively from mmWave-estimated vibration signals, enabling broader applicability and easier deployment.  Moreover, our method produces more intelligible and natural-sounding speech (\Section{overall_performance}), demonstrating superior performance compared to existing methods.

%% file: Body/background.tex
\section{Preliminary} \label{sec:pre}

\subsection{Frequency Range vs Speech Intelligibility}\label{sec:frequency_vs_intelligibility}
Speech is produced through the coordinated effort of the lungs, vocal folds, and articulators—such as the tongue, lips, and soft palate—which work together to shape airflow into meaningful sound. In non-tonal languages like English, speech intelligibility primarily relies on the clear articulation of vowels and consonants.
Interestingly, although speech spans a broad range of audible frequencies, studies have shown that applying a low-pass filter to remove frequencies above 4 kHz results in only a minor loss in intelligibility—approximately 10\% \cite{speech_intelligibility}. 
This is because the most crucial information for understanding speech lies between 500 Hz and 4 kHz, particularly around 2 kHz, where key consonants like /k/, /p/, /s/, and /t/ are concentrated \cite{french1947factors}.
\textit{This indicates that frequency components below 4 kHz are nearly sufficient for speech comprehension and, by extension, for eavesdropping tasks.} Previous studies \cite{wang2024vibspeech}\cite{li2022mmphone} have shown that mmWave radar-estimated vibration signals are naturally narrow-band. To achieve intelligible speech recovery within these constraints, we therefore focus on detecting and enhancing the perceptual quality of audio components below 4 kHz.

\subsection{Rationale of mmWave-based Vibration Sensing}
The mmWave radar using frequency-modulated continuous wave (FMCW) signals is widely adopted for its high sensing granularity. It detects vibrations by tracking the phase of the received intermediate-frequency (IF) signal at a specific range. The radar continuously transmits chirp signals, which form a frame, and captures their reflections from the target object.

Specifically, the resulting IF signal can be expressed as $S_{IF} = A \sin(2\pi f_{o}t+\phi_{o})$, where $f_o$ is the frequency, which is proportional to the distance between the radar and the vibrating object. To extract the vibration information, a Range-FFT is applied to $S_{IF}$ to identify the peak corresponding to the target's range bin. The vibration signal $\Delta d$ is then obtained by analyzing the phase variation $\Delta \phi$ over time. 
In general, improving vibration sensing quality requires a high sampling rate. However, the effective sampling rate is governed by the number of chirps per frame. While increasing the chirp count improves temporal resolution, it also shortens the duration of each chirp, which in turn reduces distance resolution, and thus impacts vibration sensing accuracy. 
In \Section{vibrationSensing}, we experimentally determine an optimal trade-off configuration that balances these competing factors for effective sensing.

%% file: Body/overview.tex
\section{Threat Model and Overview}\label{sec:threat_model}

\subsection{Threat Model}
\textbf{Attack Scenario.} As illustrated in \Figure{intro}, we consider a scenario in which individuals are engaged in an online meeting inside a room. A loudspeaker is used to play the remote participant's speech, which may contain sensitive or private information. Materials suitable for capturing vibrations are pre-positioned indoors. An attacker, equipped with a portable mmWave radar, attempts to eavesdrop by capturing sound-induced vibrations from the material, either from within the room or via external means (\eg, through walls or windows). The attacker's goal is to analyze the captured vibration signals and reconstruct semantically intelligible speech, enabling covert surveillance without needing direct access to the audio source.

\textbf{Assumption.} We assume the attacker has no prior knowledge of the victim's identity or speech characteristics. Specifically, the attacker does not have victim-specific vibration or audio samples for training the speech reconstruction DNN. Additionally, we place no restrictions on the content of the speech—meaning the victims are assumed to speak naturally, and their speech is not part of the training dataset.

\begin{figure}[t]
\centering
\begin{minipage}[t]{\linewidth}
\centering
{\includegraphics[width=.85\textwidth]{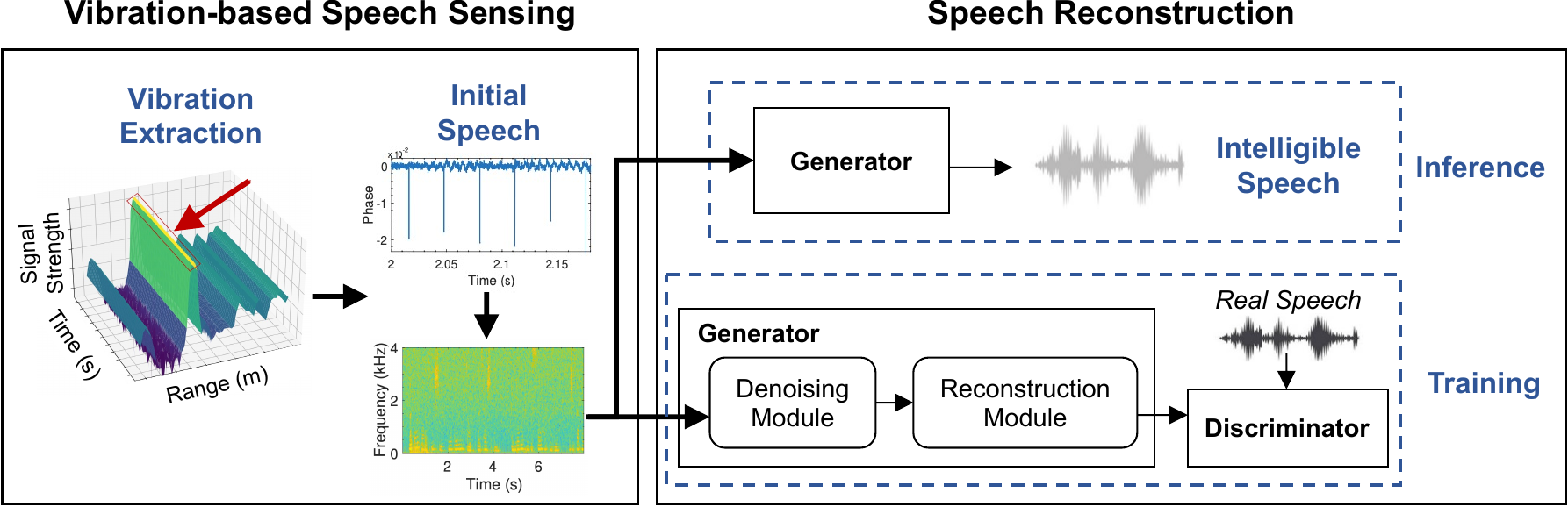}}
\caption{The \system system mainly consists of two main components, \ie, mmWave-based vibration sensing and DNN-based speech reconstruction.}
\label{fig:overview}
\end{minipage}
\vspace{-0.1in}
\end{figure}

\subsection{Overview}
As illustrated in \Figure{overview}, the \system system is composed of two key components. The first, detailed in \Section{sensing}, focuses on vibration-based speech sensing, aiming to extract high-quality vibration signals and produce an initial speech estimate. The second component, described in \Section{model}, is an end-to-end speech reconstruction network that denoises the initial speech, restores missing frequency components, and generates clear, intelligible speech. These two stages work together to enable effective speech recovery from mmWave-sensed vibrations.

%% file: Body/sensing.tex
\section{Vibration-based Speech Sensing}\label{sec:sensing}

This section outlines the core components of our vibration-based speech sensing pipeline. We start by analyzing the frequency response characteristics of different materials and discuss how to select those that enhance sensing precision. Next, we detail the signal processing pipeline used to extract vibration signals from raw mmWave radar data. We also investigate the influence of sampling rate on the quality of captured vibration signals to further optimize speech sensing quality.

\subsection{Frequency Response Modeling}\label{sec:frequency_response_modeling}

The vibration of an object caused by sound from a loudspeaker can be modeled as a damped forced vibration system—specifically, a classic mass-spring-damper system \cite{xu2020touchpass}. Given an external force $F_0$ (generated by the loudspeaker), the frequency response of the resulting vibration can be expressed as:
\begin{equation}\label{eq:freq_response}
X(w) = \frac{F_0}{\sqrt{(k-mw^2)^2+(cw)^2}}.
\end{equation}
Here, $X(w)$ represents the vibration amplitude of the object at frequency $w$, and the denominator captures the system's impedance characteristics. The parameters $k$, $m$, $c$ denote the stiffness, mass, and damping coefficient of the vibrating object, respectively.

This model reveals several key behaviors of the system:
(1) When $w$ equals the vibrating object's natural frequency $w_n = \sqrt{k/m}$, the vibration amplitude $X(w)$ reaches its maximum.
(2) For frequencies $w>w_n$, the vibration amplitude rapidly decreases, with the rate of decay influenced by the object's mass $m$.
In practice, this means that higher-frequency vibrations are weaker and more prone to noise, limiting the ability to capture fine speech details. Therefore, the physical properties of the vibrating object—especially its mass and natural
frequency—are critical for effective vibration-based sensing. To preserve a broader frequency range, particularly the high-frequency components vital for intelligible speech, materials with higher natural frequencies $w_n$ and lower mass $m$ are preferable. This theoretical insight is validated by our experimental results in \Section{vibration_sensing}.

\begin{figure*}[t]
\centering
\begin{minipage}[t]{\textwidth}
\centering
\subfigure[Beginning outlier]
{\includegraphics[width=.245\textwidth]{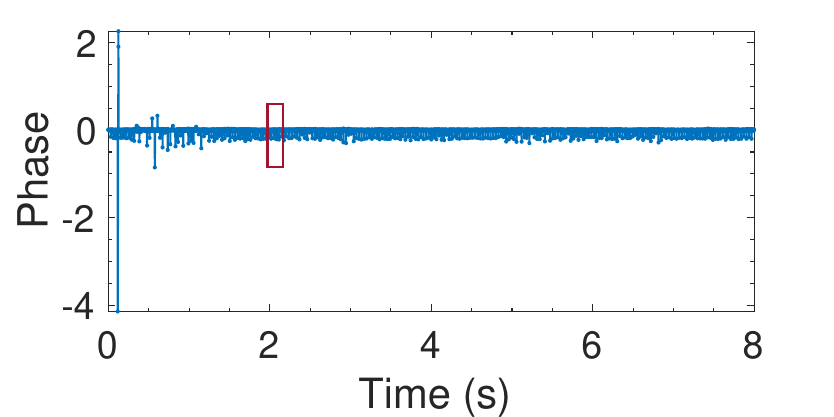}}
% \hspace{0.1in}
\subfigure[Periodic outlier]
{\includegraphics[width=.245\textwidth]{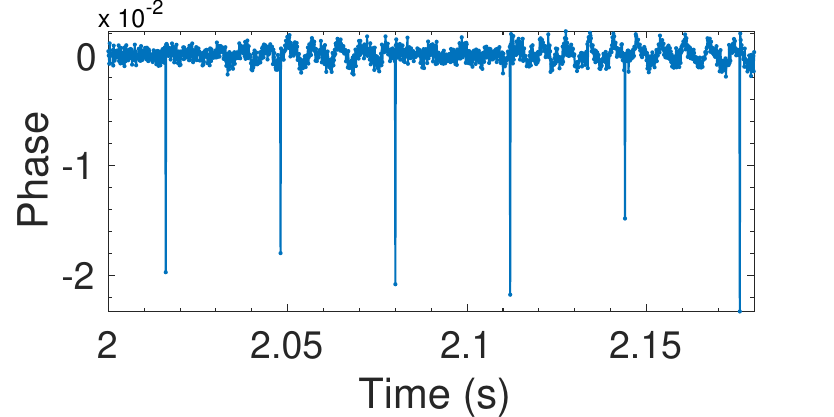}}
\subfigure[After preprocessing]
{\includegraphics[width=.245\textwidth]{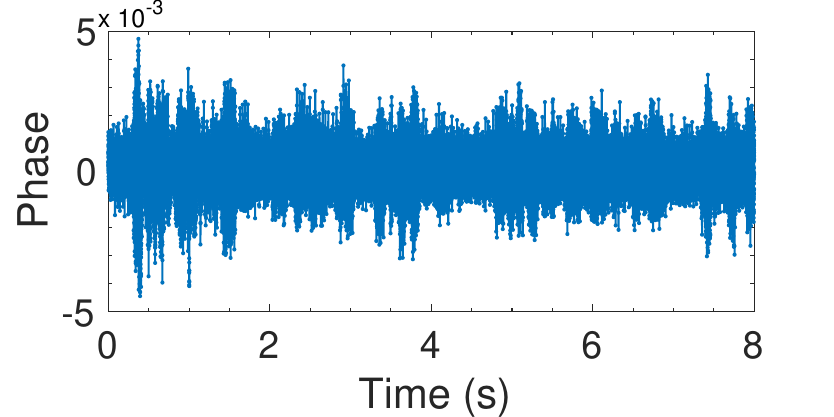}}
% \hspace{0.1in}
\subfigure[The played speech]
{\includegraphics[width=.245\textwidth]{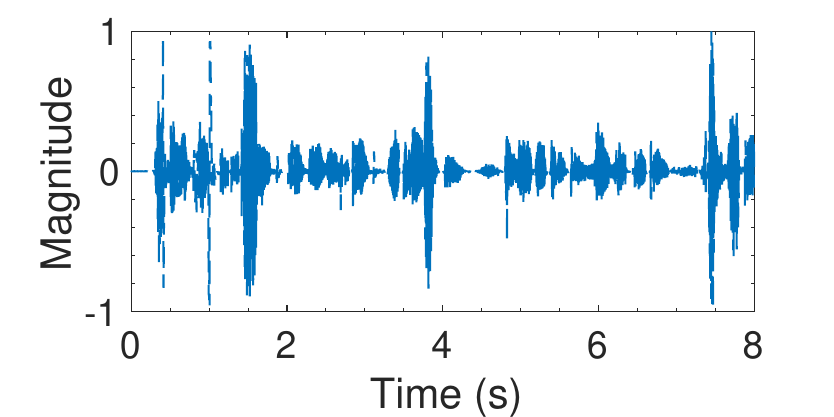}}
\subfigure[Spectrogram of (a)]
{\includegraphics[width=.245\textwidth]{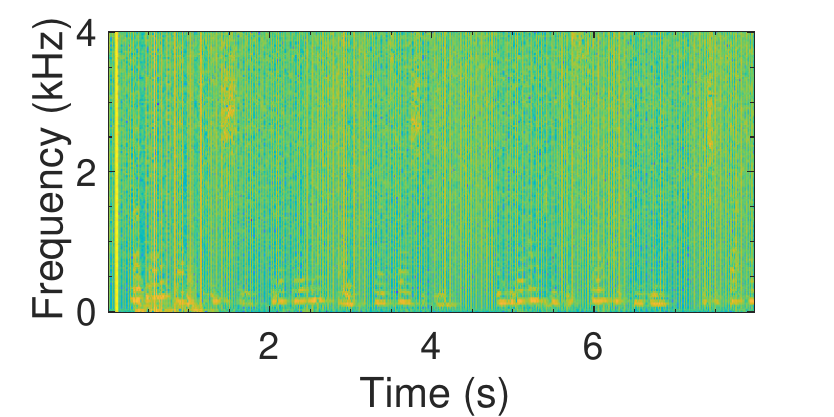}}
% \hspace{0.1in}
\subfigure[Spectrogram of (b)]
{\includegraphics[width=.245\textwidth]{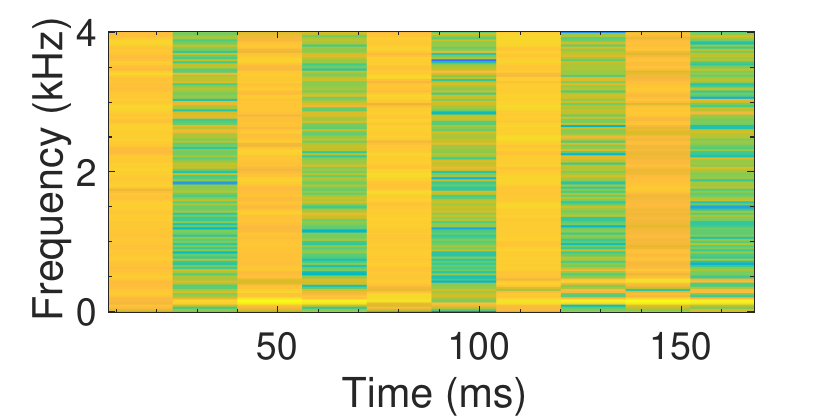}}
\subfigure[Spectrogram of (c)]
{\includegraphics[width=.245\textwidth]{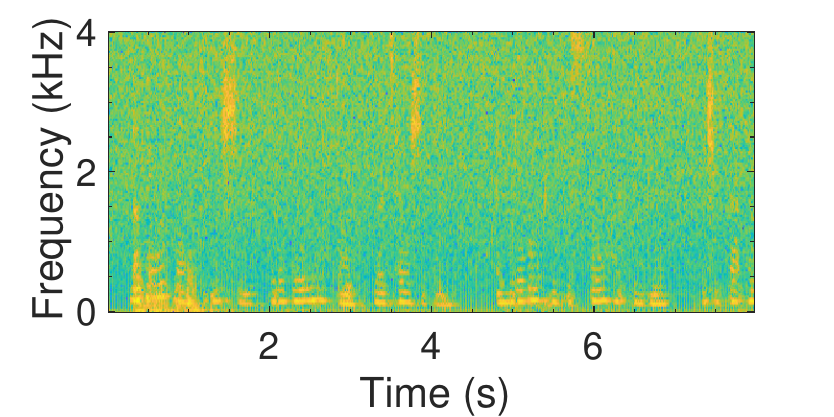}}
% \hspace{0.1in}
\subfigure[Spectrogram of (d)]
{\includegraphics[width=.245\textwidth]{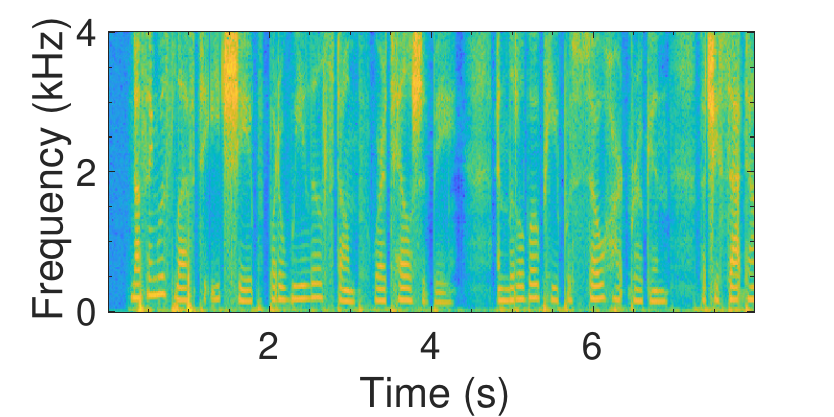}}
\vspace{-0.1in}
\caption{Preprocessing of the mmWave radar estimated vibration signals.}
\label{fig:de_preprocess}
\end{minipage}
\vspace{-0.1in}
\end{figure*}

\subsection{mmWave-based Vibration Extraction}\label{sec:vibrationSensing}

To investigate the capability of mmWave radar in sensing sound-induced vibrations, we conducted proof-of-concept experiments using various vibrating objects composed of different materials. The test setup included a JBL loudspeaker and two passive films: tinfoil (commonly used in prior studies \cite{wang2024vibspeech}\cite{wang2024vibration}) and aluminum-coated PET (polyethylene terephthalate), hereafter referred to as PET. These objects simulated both direct eavesdropping from the loudspeaker surface and indirect eavesdropping via passive films. 
PET, in particular, is a widely used material found in everyday items such as food packaging (\eg, snack bags, coffee pouches) and thermal insulation products (\eg, space blankets, outdoor blanket).

During the experiment, we played randomly selected speech samples from the LibriSpeech dataset \cite{panayotov2015librispeech} through the JBL loudspeaker at a sound pressure level (SPL) of 75 dB. An mmWave radar, positioned 1.5 m in front of each test object, transmitted chirps in the 60–64 GHz range to capture the resulting vibrations. Specifically, for the passive film tests, the material was placed between the radar and the loudspeaker at a distance of 0.5 m from the sound source.

\subsubsection{Preprocessing}
To extract sound-induced vibrations, we first apply Range-FFT to the collected IF signals from the radar to identify the target range bin with the strongest reflection. We then track phase variations within this bin over time to capture the vibration signal. During this process, we observe two types of outliers that degrade the signal-to-noise ratio (SNR) in both time and frequency domains. 
As illustrated in \Figure{de_preprocess}(a), a prominent spike appears at the beginning of each recording—referred to as the \textit{beginning outlier}, whose value is significantly larger than those in subsequent frames. 
Additionally, the radar transmits chirps in a discontinuous manner, \ie, there are inter-frame gaps during which no chirps are transmitted. This discontinuity can cause power fluctuations within the radar hardware. As a result, when zooming in on the signal (\Figure{de_preprocess}(b)), periodic spikes appear at the start of each frame—referred to as \textit{periodic outliers}.

Since the vibrations induced by the loudspeaker are of very low magnitude, these outliers can easily overwhelm the actual vibration patterns, as demonstrated in \Figure{de_preprocess}(e)(f)). To address these artifacts, we apply a two-stage preprocessing method. First, we remove the beginning outlier using the 3-sigma rule: any data point deviating from the mean by more than three standard deviations is replaced with the mean value. Second, we eliminate periodic outliers by leveraging their fixed positions (based on the known sampling rate) and replacing those values with the local mean.
The result after preprocessing is shown in \Figure{de_preprocess}(c). Comparing its spectrogram (\Figure{de_preprocess}(g)) with \Figure{de_preprocess}(e)(f) as well as that of the original audio (\Figure{de_preprocess}(h)), we observe that the primary vocal patterns are successfully revealed.

\begin{figure*}[t]
\centering
\begin{minipage}[t]{\textwidth}
\centering
\subfigure[Audio]
{\includegraphics[width=.195\textwidth]{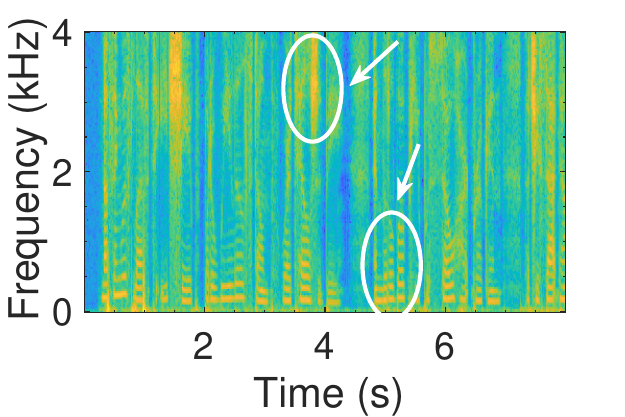}}
% \hspace{0.1in}
\subfigure[Loudspeaker]
{\includegraphics[width=.195\textwidth]{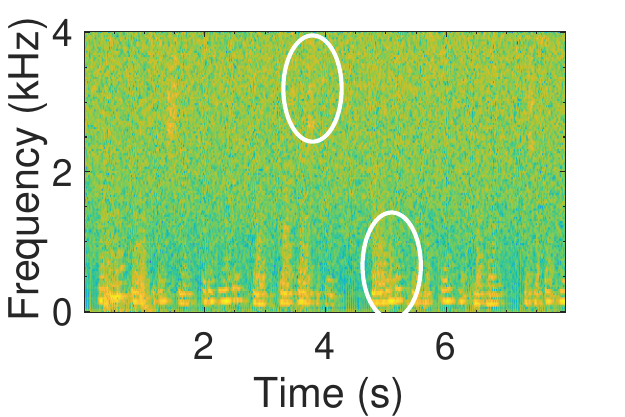}}
\subfigure[Tinfoil]
{\includegraphics[width=.195\textwidth]{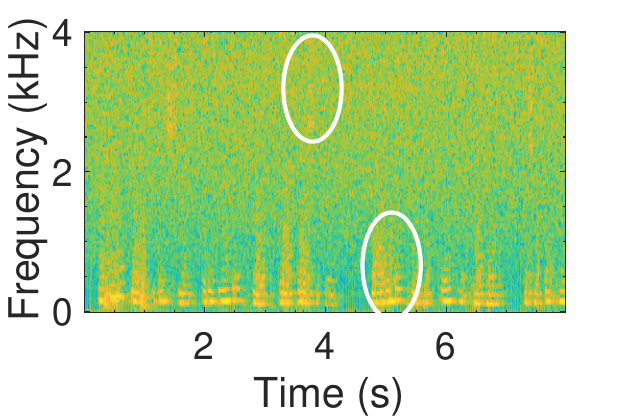}}
% \hspace{0.1in}
\subfigure[PET]
{\includegraphics[width=.195\textwidth]{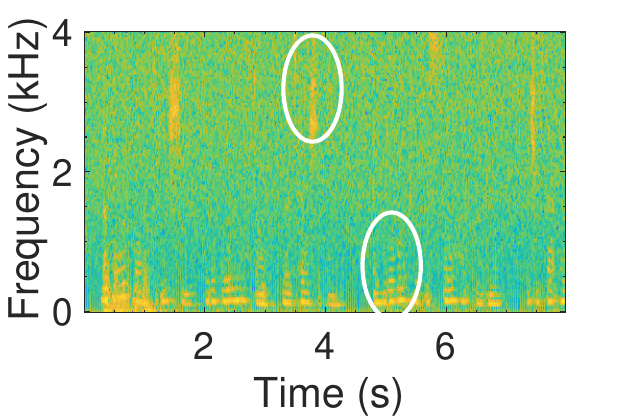}}
\subfigure[PET-Through wall]
{\includegraphics[width=.195\textwidth]{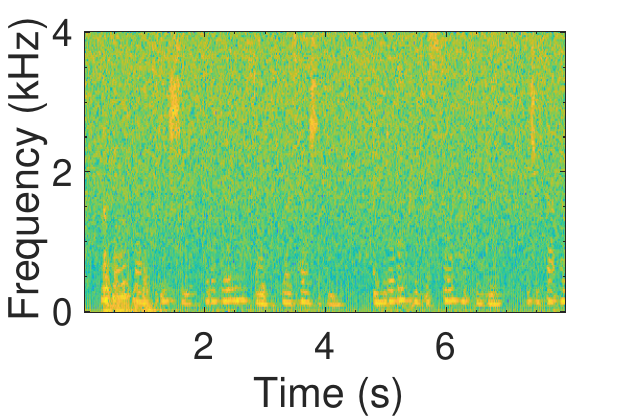}}
\vspace{-0.1in}
\caption{Frequency response of different vibrating materials.}
\label{fig:de_vib}
\end{minipage}
\vspace{-0.1in}
\end{figure*}

\begin{figure*}[t]
\centering
\begin{minipage}[t]{\textwidth}
\centering
\subfigure[PET-16000Hz]
{\includegraphics[width=.24\textwidth]{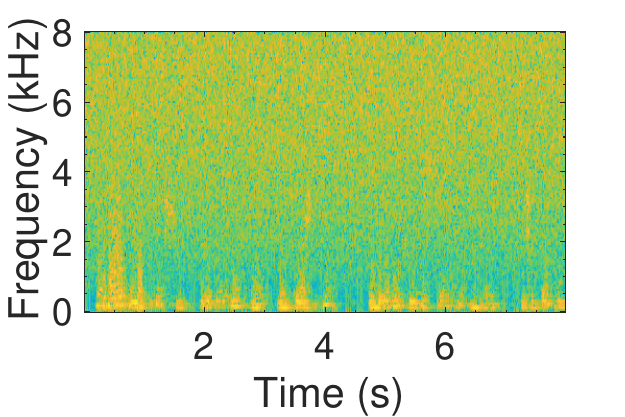}}
% \hspace{0.1in}
\subfigure[Audio]
{\includegraphics[width=.24\textwidth]{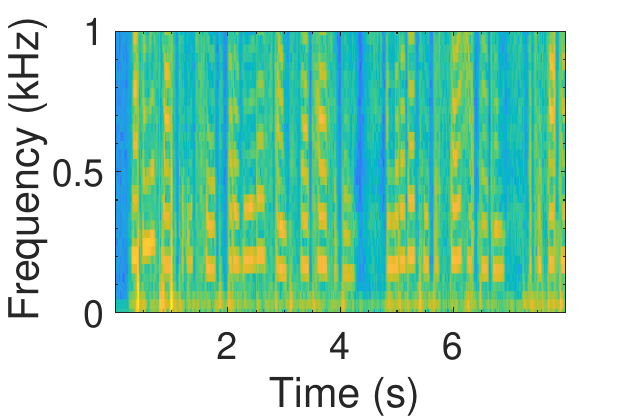}}
\subfigure[PET-8000Hz]
{\includegraphics[width=.24\textwidth]{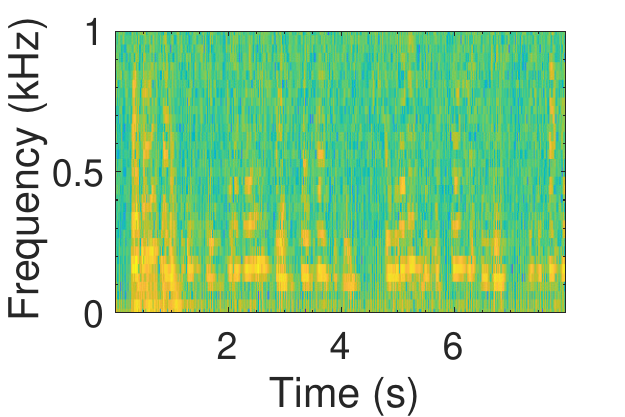}}
% \hspace{0.1in}
\subfigure[PET-16000Hz]
{\includegraphics[width=.24\textwidth]{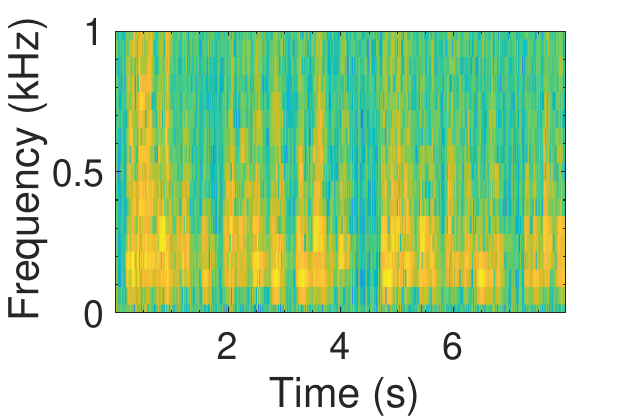}}
\vspace{-0.1in}
\caption{Frequency response vs sampling rate.}
\label{fig:de_samplingRate}
\end{minipage}
\vspace{-0.1in}
\end{figure*}

\subsubsection{Vibration Sensing}\label{sec:vibration_sensing}
As discussed in \Section{frequency_response_modeling}, material properties significantly influence vibration sensing bandwidth. To evaluate this effect, we played the same speech sentence by the JBL loudspeaker and measured the resulting vibrations using the previously mentioned materials. The original speech spectrum and each material's vibration response are presented in \Figure{de_vib}.

\textit{Vibration Property.}
As shown in \Figure{de_vib}(a), most speech energy is concentrated below 1 kHz, and all three materials successfully captured these low-frequency components. However, when examining details such as the vocal pattern around the fifth second, the signals from the loudspeaker and tinfoil appear noticeably degraded compared to the original speech, whereas PET retains clearer features. Additionally, the vibration amplitudes of all materials decline with increasing frequency, consistent with the prediction in \Equation{freq_response}. Among them, PET demonstrates superior preservation of high-frequency vocal features, attributed to its favorable material properties. 
This result confirms to our analysis introduced in \Section{frequency_response_modeling}. PET offers a high natural frequency—reaching several kHz—and has only one-third the mass of tinfoil film. Its aluminized surface also provides strong reflectivity for mmWave signals. These factors make it a more suitable material for our mission.
Therefore, we select PET as the default sensing medium for subsequent speech recovery experiments.

\textit{Sampling Rate Selection.} 
The sampling rate of the radar also affects its vibration sensing performance. We tested two rates—8000 Hz and 16000 Hz—corresponding to theoretical upper capture limits of 4 kHz and 8 kHz, respectively. Spectrograms of the resulting signals are shown in \Figure{de_samplingRate}. At 16000 Hz (\Figure{de_samplingRate}(a)), significant high-frequency information (>4 kHz) is lost, aligning with our earlier analysis that mmWave-based vibration sensing struggles to retain high-bandwidth content due to noise.
Interestingly, increasing the sampling rate to 16000 Hz also cause the low-frequency components to become blurred (\Figure{de_samplingRate}(b)(c)(d)). This is because a higher sampling rate requires more chirps per frame, which in turn shortens the duration of each chirp. Shorter chirps reduce distance resolution, thereby degrading the quality of vibration sensing.

\textit{Through-Wall Sensing.} 
To assess the radar's through-wall eavesdropping capability, we placed it behind a glass wall and measured the vibration response (\Figure{de_vib}(e)) with the sampling rate of 8000 Hz. Compared with the direct PET response (\Figure{de_vib}(d)), the low-frequency components remain detectable. However, the high-frequency band suffers from increased background noise, though it is still more distinguishable than the response of tinfoil in \Figure{de_vib}(c). As discussed in \Section{training_process}, this issue can be mitigated using our proposed noise-based signal synthesis method. Specifically, our speech reconstruction DNN model will be pre-trained with synthesized signals containing higher levels of noise to improve performance.

\textbf{Summary.} To capture fine-grained vibrations, the chosen vibration-sensitive surface should exhibit a high natural frequency, low mass, and strong mmWave reflectivity. Additionally, both the radar's sampling rate and its distance resolution play critical roles in determining vibration sensing quality. Based on these considerations—and informed by findings in \Section{frequency_vs_intelligibility} that limiting speech to frequencies below 4 kHz leads to only a 10\% loss in intelligibility—we select an 8000 Hz radar sampling rate. This setting provides a practical trade-off between frequency coverage and sensing resolution when using PET, making it well-suited for effective speech recovery and through-wall eavesdropping.

%% file: Body/model.tex
\section{Speech Reconstruction Model for Refinement}\label{sec:model}

\begin{figure}[t]
\centering
\begin{minipage}[t]{.85\linewidth}
{\includegraphics[width=\textwidth]{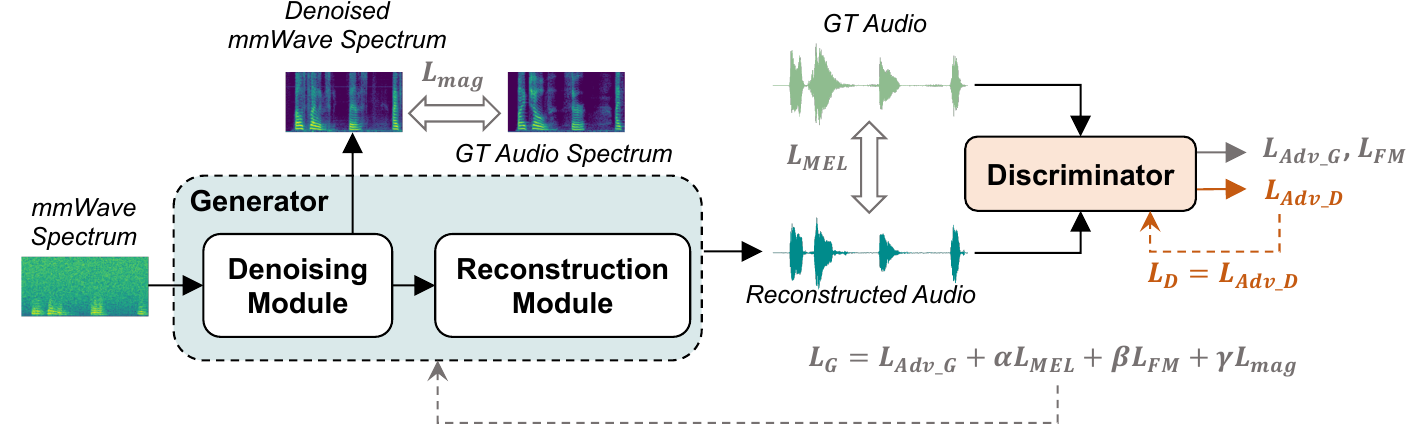}}
\caption{The overall speech reconstruction DNN model of \system.}
\label{fig:mod_model}
\end{minipage}
\vspace{-0.1in}
\end{figure}

Building on the initial speech signal derived from mmWave-based vibration sensing (\Section{vibrationSensing}), we then develop an end-to-end deep neural network (DNN) to enhance speech quality. 
As illustrated in \Figure{mod_model}, the architecture is based on a Generative Adversarial Network (GAN), commonly used in speech synthesis not as pure generators, but as perceptual enhancers.
Our DNN consists of two main components: a \textit{Generator} and a \textit{Discriminator}.The Generator processes the spectrogram of the input vibration signal, first passing it through a \textit{Spectrum Denoising} module to suppress noise. The denoised spectrogram is then fed into a \textit{Speech Reconstruction} module, which restores missing frequency components and converts the result into a time-domain waveform.
To further refine perceptual quality, a hybrid Discriminator evaluates the reconstructed speech. Using an adversarial learning strategy, the Discriminator provides feedback that enables the Generator to iteratively produce speech with improved intelligibility and naturalness for human listeners.

\subsection{Generator}
\subsubsection{Spectrum Denoising}

\begin{figure}[t]
\centering
\begin{minipage}[t]{.9\linewidth}
\centering
\subfigure[Denoising module]
{\includegraphics[width=.2\textwidth]{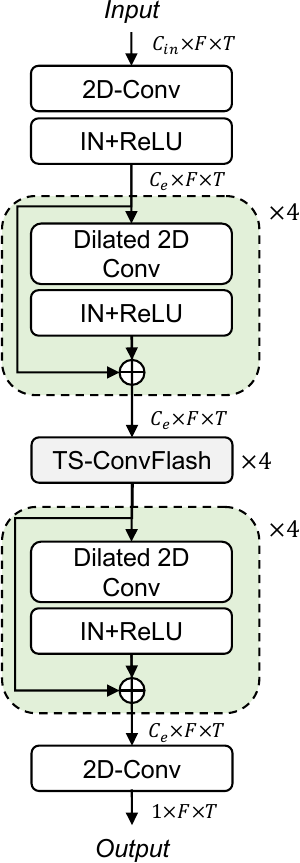}}
\hspace{0.35in}
\subfigure[TS-ConvFlash block]
{\includegraphics[width=.55\textwidth]{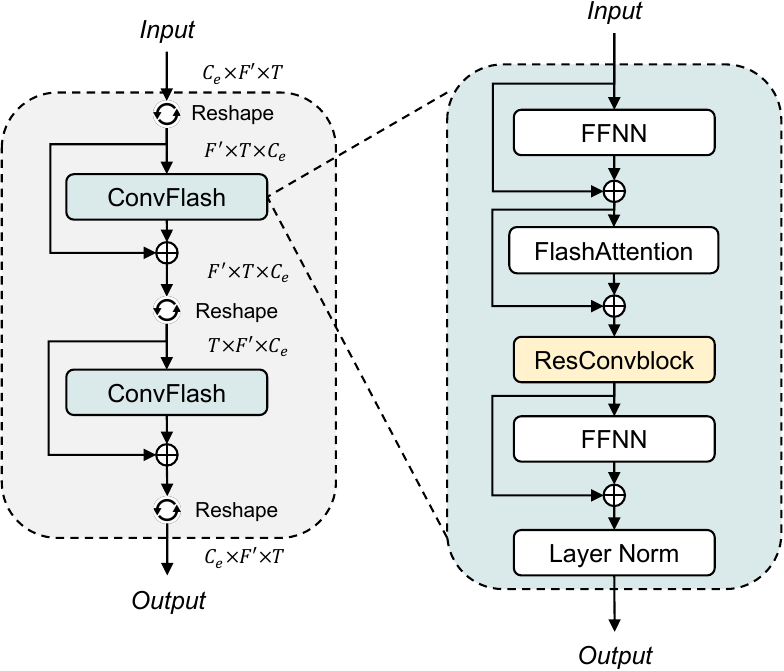}}
\vspace{-0.1in}
\caption{The structure of spectrum denoising module.}
\label{fig:mod_denoising}
\end{minipage}
\vspace{-0.1in}
\end{figure}

As illustrated in \Figure{mod_denoising}(a), the spectrum denoising module takes the raw mmWave spectrogram $X \in \mathbb{R}^{C_{in}\times F\times T}$ as input, where $C_{in}$ is the number of input channels, $F$ is the number of frequency bins, and $T$ is the temporal length. This input is first processed through a 2D convolution layer, followed by Instance Normalization and a ReLU activation, expanding the channel dimension to $C_e$ and producing an initial feature map $x_{enc}$.
To capture multi-scale features, $x_{enc}$ is then passed through $N$ layers of dilated convolutions with varying dilation rates. The output of these layers, $H_N$, is then fed into our proposed TS-ConvFlash (\Figure{mod_denoising}(b)), which captures temporal and frequency-domain dependencies, an enhanced version of the TS-Conformer from CMGAN \cite{cao2022cmgan}. By replacing the original attention mechanism with FlashAttention \cite{dao2022flashattention}, TS-ConvFlash maintains sequence modeling capabilities while significantly improving efficiency. It reshapes the time and frequency dimensions to independently model temporal and spectral dependencies, generating refined feature representations $A_{out}$. These features are further processed by additional dilated convolutions, and finally, a 2D convolution reduces the channel dimension to 1, yielding the denoised spectrogram $X_{mag}\in \mathbb{R}^{1\times F\times T}$.

\subsubsection{Speech Reconstruction}
\begin{figure}[t]
\centering
\begin{minipage}[t]{.9\linewidth}
\centering
\subfigure[Reconstruction module]
{\includegraphics[width=.55\textwidth]{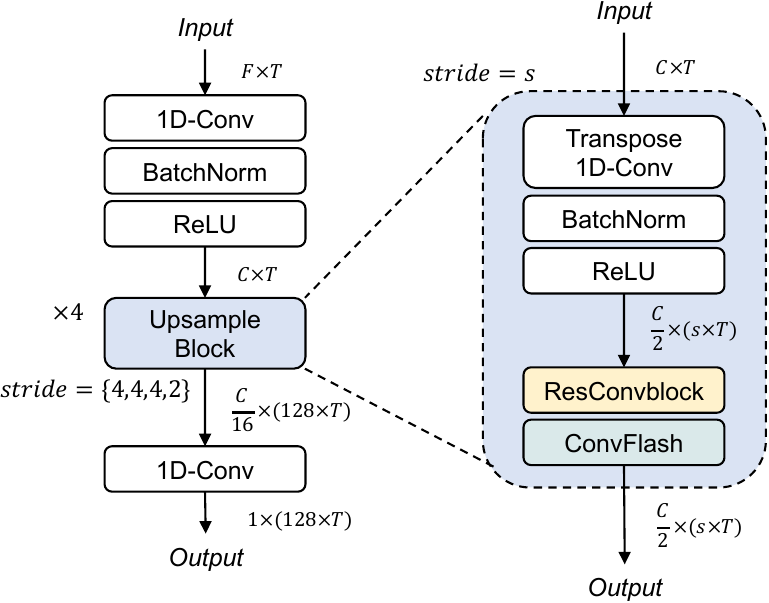}}
\hspace{0.35in}
\subfigure[ResConv block]
{\includegraphics[width=.32\textwidth]{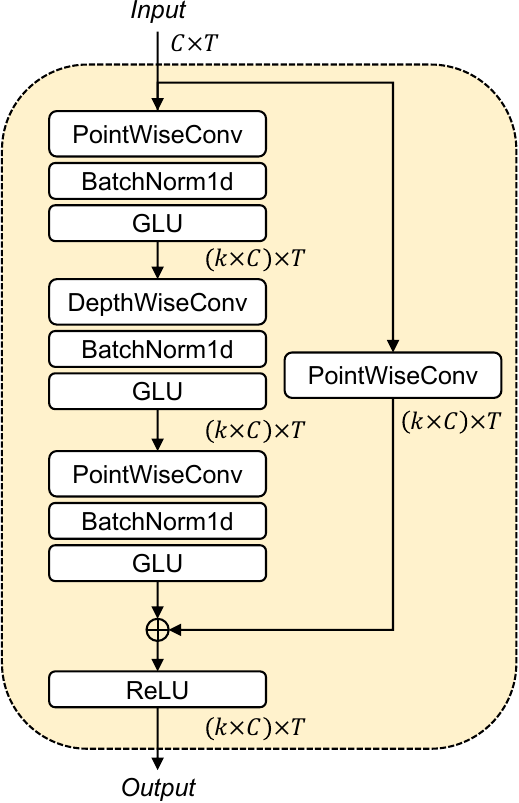}}
\vspace{-0.1in}
\caption{The structure of speech reconstruction module.}
\label{fig:mod_reconstructionModule}
\end{minipage}
\vspace{-0.15in}
\end{figure}

As shown in \Figure{mod_reconstructionModule}, the denoised spectrogram $X_{mag}$ is then processed by the Speech Reconstruction module, which generates the final time-domain speech signal $X_{rebuild}\in \mathbb{R}^{1\times T_{rebuild}}$.
In detail, the reconstruction process begins with a 1D convolution followed by Batch Normalization and ReLU activation function, expanding the channel dimension $F$ to $C$, producing an intermediate feature $X'_{mag}$.
This feature is then progressively upsampled using a series of UpsampleBlocks. Each block upsamples the input by a factor of $s$ using transposed 1D convolution, reduces the channel dimension to $C/2$ via a ResConvBlock (\Figure{mod_reconstructionModule}(b)), and applies a ConvFlash module for temporal sequence modeling. 
After four UpsampleBlocks with upsampling factors $s = [4,4,4,2]$, the temporal resolution of $X'_{mag}$ increases by a factor of 128, while the channel dimension returns to $C$. Finally, a 1D convolution reduces the channel count to 1, producing the reconstructed waveform $X_{rebuild}\in \mathbb{R}^{1\times 128\times T}$.

\subsection{Discriminator}\label{sec:discriminator}

To ensure the reconstructed speech aligns with human auditory perception, we adopt adversarial training, a common practice in recent speech synthesis methods. 
Inspired by BigVGAN \cite{lee2022bigvgan}, our discriminator combines a Multi-Period Discriminator (MPD) \cite{kong2020hifi} with a Multi-Resolution Discriminator (MRD) \cite{you2021gan} to jointly suppress periodic artifacts and enhance perceptual quality.

The MPD consists of multiple sub-discriminators, each sampling audio at different intervals (2, 3, 5, 7, 11). This configuration minimizes receptive field overlap and enables each sub-discriminator to focus on specific periodic patterns, such as pitch and harmonics.
The MRD, on the other hand, also comprises multiple sub-discriminators. MRD is responsible to downsample the waveform into various resolutions and computes the Short-Time Fourier Transform (STFT) spectrogram for each. High-resolution sub-discriminators focus on capturing local details like high-frequency noise and transient phase variations, while low-resolution sub-discriminators attend to global structure and long-term dependencies such as fundamental frequency and prosody. 
Together, these discriminators guide the generator to generate speech with fine-grained details and enhanced frequency components, pushing the generated audio to be realistic across both temporal and spectral domains.

\subsection{Loss Function}\label{sec:loss}
To optimize both the generator and discriminator, we introduce four loss functions.

\subsubsection{Spectrum Denoising Loss}
This loss optimizes the spectrum denoising module by minimizing the L1 distance between the output spectrogram ($x_{mag}$) of the module and the ground truth speech spectrogram ($s_{mag}$). It is defined as:
\begin{equation}\label{eq:L_denoise}
L_{mag} = \mathbb{E} \begin{Vmatrix}
x_{mag} - s_{mag}
\end{Vmatrix}_1.
\end{equation}

\subsubsection{MEL Loss}

To enhance frequency modeling across different time scales, we adopt a multi-resolution Mel-spectral loss (short for MEL loss $L_{MEL}$):
\begin{equation}\label{eq:L_denoise}
L_{MEL} = \sum_{j=1}^{7} \mathbb{E} \begin{Vmatrix} \varphi_j(s)-\varphi_j(G(x)) \end{Vmatrix}_1,
\end{equation}
where $\varphi(\cdot)$ denotes the Mel spectrogram computation function for the $j$-th group of setting. In our implementation, we use seven different sets of Mel spectrogram parameters to capture features at multiple resolutions (see \Section{mel_loss} for details).

\subsubsection{Adversarial Loss}
As introduced in \Section{discriminator}, we employ two discriminators—MPD and MRD. MPD consists of five sub-discriminators, while MRD comprises four. The adversarial training loss is defined as follows:
\begin{equation}\label{eq:L_adv}
\begin{split}
L_{Adv\_D} &= \sum_{i=1}^{2}\sum_{k=1}^{K_i} \mathbb{E}_{(x,s)} 
\begin{bmatrix}
(1-D_{i,k}(s))^2 + (D_{i,k}(G(x)))^2
\end{bmatrix},\\
L_{Adv\_G} &= \sum_{i=1}^{2}\sum_{k=1}^{K_i}  \mathbb{E}_x \begin{bmatrix}
(1-D_{i,k}(G(x)))^2
\end{bmatrix}.
\end{split}
\end{equation}
Here, $D_{i,k}$ represents $k$-th sub-discriminator of the $i$-th discriminator (\ie, MPD or MRD), and $K_i$ denotes the number of sub-discriminators within the $i$-th discriminator. $s$ refers to real speech data, while $x$ denotes the generated speech from mmWave signals.
This group of adversarial loss encourages the generator to produce outputs that are indistinguishable from real speech, enhancing the naturalness of the reconstructed speech.

\subsubsection{Feature Matching Loss}

The feature matching loss is a similarity metric based on differences in feature representations extracted by the discriminator, which is defined as:
\begin{equation}\label{eq:L_FM}
L_{FM} = \mathbb{E}_{(x,s)} \begin{bmatrix} \sum_{i=1}^{T} \frac{1}{N_i} \begin{Vmatrix} D_i(s)-D_i(G(x)) \end{Vmatrix}_1
\end{bmatrix},
\end{equation}
where $T$ is the number of layers in the discriminator, $N_i$ is the dimension of features at the $i$-th layer, and $D_i(s)$ and $D_i(G(x))$ refer to the feature representations of the real and generated speech, respectively, at layer $i$ (see \Section{feature_loss} for details).

\textbf{Summary.} 
The total loss of our network is:
\begin{equation}
\begin{split}
L_G &= L_{Adv\_G}+ \lambda_{MEL} L_{MEL} +  \lambda_{FM} L_{FM} + \lambda_{mag} L_{mag}\\
L_D &= L_{Adv\_D}
\end{split}
\end{equation} 
As shown in \Figure{mod_model}, $L_G$, $L_D$ is used for training the generator and discriminator, respectively. In the implementation, we set $\lambda_{MEL}=7$, $\lambda_{FM}=1.5$, and $\lambda_{mag}=2$.

\section{mmWave-based Vibration Data Synthesis}\label{sec:training_process}

The performance and generalization ability of deep learning models depend on the quantity and diversity of training data. However, real-world mmWave vibration datasets are scarce. To address this limitation, we propose a data augmentation strategy that synthesizes mmWave-like signals using clean speech samples from the widely adopted LibriSpeech corpus \cite{panayotov2015librispeech}\cite{ding2022ultraspeech}.

Our observation of real mmWave-based vibration signals reveals a frequency-dependent noise pattern: specifically, high-frequency components tend to exhibit characteristics of purple (or violet) noise, where power density increases with frequency. To replicate this degradation, we introduce a mixture of purple noise and Gaussian noise with randomly scaled intensities to clean speech signals. This approach simulates the spectrum distortion found in mmWave data. 
The synthesized signal is computed as:
\begin{equation}\label{eq:noise_generation}
X_{mmVib} = Nom(X_{speech})+\alpha N_{pruple}+\beta N_{Gaussian}.
\end{equation}
Here, $X_{speech}$ is the clean input speech, while $N_{pruple}$ and $N_{Gaussian}$ are purple and Gaussian noise components with unit variance. $Nom(\cdot)$ denotes the Z-score operation for normalization, and we set $\alpha=1$  and $\beta=0.3$.

\begin{figure*}[t]
\centering
\begin{minipage}[t]{\textwidth}
\centering
\subfigure[Audio]
{\includegraphics[width=.195\textwidth]{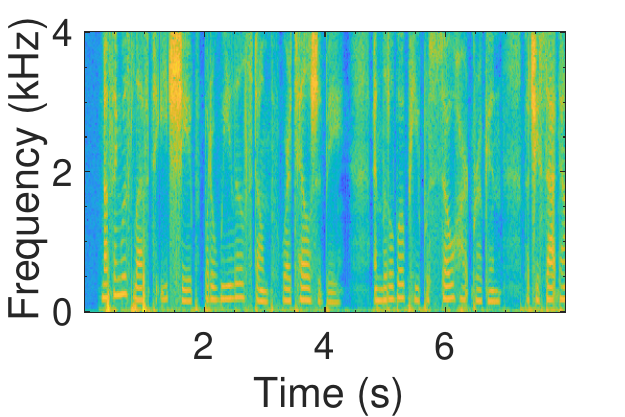}}
% \hspace{0.1in}
\subfigure[PET]
{\includegraphics[width=.195\textwidth]{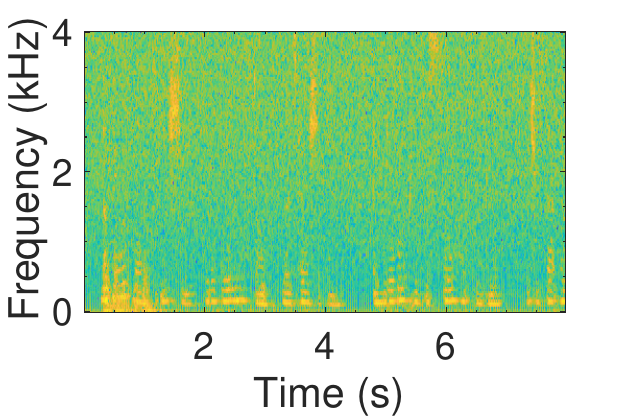}}
\subfigure[Audio \& Purple Noise]
{\includegraphics[width=.195\textwidth]{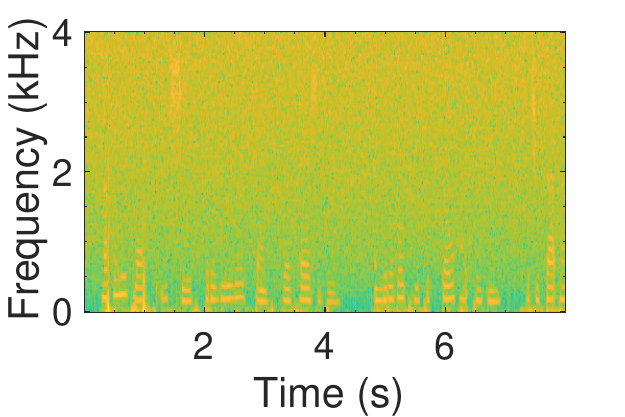}}
% \hspace{0.1in}
\subfigure[Audio \& Gaussian Noise]
{\includegraphics[width=.195\textwidth]{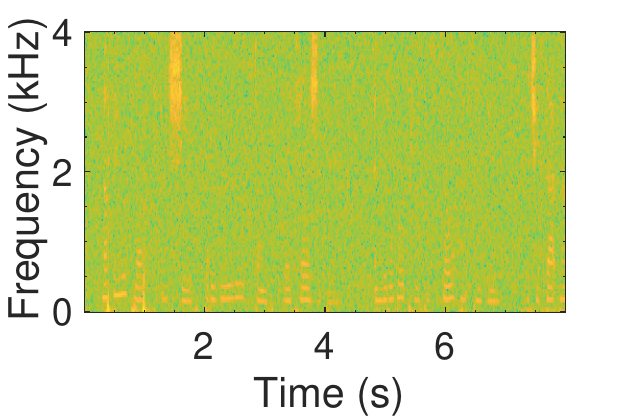}}
\subfigure[Audio \& Mixed Noise]
{\includegraphics[width=.195\textwidth]{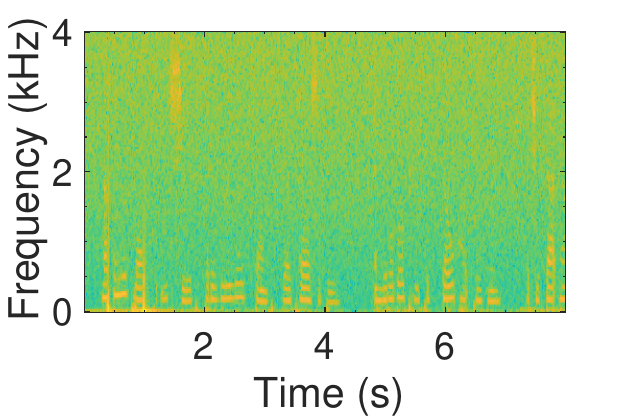}}
\vspace{-0.1in}
\caption{Spectrograms of original speech, real mmWave vibration signals, and synthesized mmWave signals for comparison.}
\label{fig:mod_noise}
\end{minipage}
\vspace{-0.1in}
\end{figure*}

\Figure{mod_noise} compares the synthesized spectrograms with those of original speech and real mmWave vibration signals. As shown in \Figure{mod_noise}(c), purple noise effectively attenuates high-frequency content, mimicking how physical vibrations decay with increasing frequency. Meanwhile, Gaussian noise (\Figure{mod_noise}(d)) simulates lower-frequency noise typically found in mmWave measurements.
When combined, as in \Figure{mod_noise}(e), the synthesized spectrogram closely resembles that of real mmWave data captured via PET (\Figure{mod_noise}(b)), validating the effectiveness of our synthetic signals.
These generated samples can simulate a wide range of sensing conditions and speaker variations, each exhibiting different levels of high-frequency attenuation.
Leveraging this capability, we adopt a two-stage training strategy to improve model robustness. We first pre-train our DNN using the synthesized signals, then fine-tune it with real mmWave measurements collected during our experiments, ensuring better generalization.

%% file: Body/implementation.tex
\section{Implementation}\label{sec:implementation}

\subsection{Experimental Setups}

\begin{figure}[t]
\centering
\begin{minipage}[t]{.6\linewidth}
{\includegraphics[width=\textwidth]{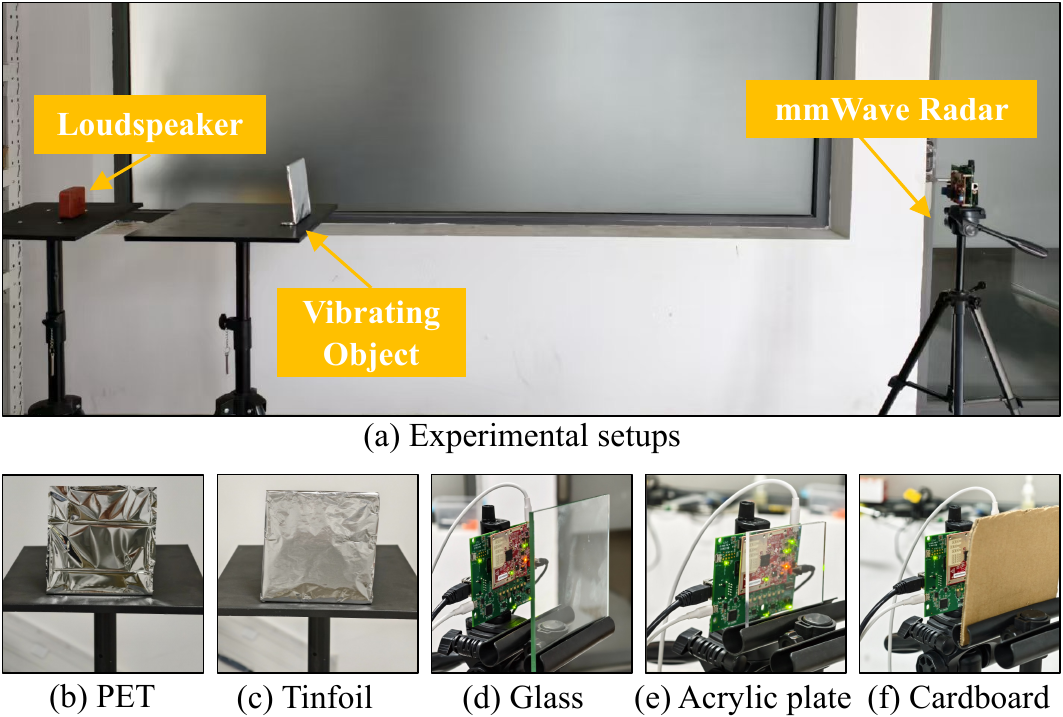}}
\caption{Experimental setups and various materials.}
\label{fig:scene}
\end{minipage}
\vspace{-0.15in}
\end{figure}

We use the commercial TI IWR6843ISK mmWave radar jointly with the DCA1000EVM for mmWave signal collection. The radar is equipped with 3 Tx antennas and 4 Rx antennas, operating at a 4GHz bandwidth. During data collection, a 1-Tx, 4-Rx (1T4R) configuration is used, with each frame lasting 32 ms and containing 256 chirps—resulting in a transmission rate of 8,000 chirps per second (\ie, vibration sampling rate of 8,000 Hz).
An illustration of the experimental setup is shown in \Figure{scene}(a). The sensing material is placed between the radar and the speaker, approximately 0.5 meters from the speaker and 1.5 meters from the radar.
The speech is played using a COTS speaker (\ie, JBL Go 2 wireless bluetooth speaker). The audio samples are sourced from the LibriSpeech dataset \cite{panayotov2015librispeech}, from which 2,400 clips—each no shorter than 8 seconds—were selected. The selected speeches includes recordings from 47 unique speakers, covering a range of genders, ages, and accents. All speech samples are distinct, with no repeated sentences across the dataset. Our experiment is conducted with approval from the Institutional Review Board (IRB).

\subsection{Dataset}\label{sec:dataset}
We construct three datasets for comprehensive evaluation.

$\mathbb{D}_{pre}$: Based on the train-clean-100 subset of the LibriSpeech corpus, a generated dataset $\mathbb{D}_{pre}$ was constructed using the method described in \Section{training_process}. This dataset contains 26,588 speech samples and is used for model pretraining. Importantly, both the speeches and speakers in $\mathbb{D}_{pre}$ are entirely distinct from the 2,400 samples used in $\mathbb{D}1$ and $\mathbb{D}2$, ensuring a clean separation for unbiased training.

$\mathbb{D}1$ (Seen Victims): This dataset is created by randomly selecting 80\% of the 2,400 collected speech samples for training, 10\% for validation, and 10\% for testing. There is no overlap in utterances across the training, validation, and test sets.

$\mathbb{D}2$ (Unseen Victims): From the collected 2,400 samples, 1,894 samples from 36 users were selected as the training set. An additional 235 samples from 4 users were used as the validation set, while 271 samples from 7 users formed the test set. This dataset configuration is designed to evaluate \system's performance across previously unseen speakers.

\subsection{DNN Implementation}

The DNN model in this work is implemented using PyTorch and trained on an NVIDIA RTX 4090 GPU. The Adam optimizer is used during training, with the learning rate adjusted via the ExponentialLR scheduler \cite{li2019exponential}, using a decay factor of 0.99996. 
For pretraining, the batch size is set to 4, and training is distributed across 4 GPUs with an initial learning rate of 6e-4 for a total of 60 epochs. In subsequent training stages, the model is trained on a single GPU with a batch size of 4, an initial learning rate of 1.5e-4, and runs for 200 epochs.

%% file: Body/evaluation.tex
\section{Evaluation}

\subsection{Metrics}
We use five widely adopted metrics to evaluate the quality of reconstructed speech:

\textbf{Frequency Weighted Segmental SNR (FWSegSNR)}: This metric evaluates the similarity between the reconstructed and the grouth truth speech by calculating their signal-to-noise ratio (SNR) in the frequency domain. It scores range from negative infinity to positive infinity, with higher values indicating better quality.

\textbf{Short-Time Objective Intelligibility (STOI)}: STOI predicts the intelligibility of speech by analyzing its time-frequency characteristics. Scores range from 0 to 1, with higher values indicating greater intelligibility.

\textbf{Mel-Cepstral Distortion (MCD)}: MCD quantifies the level of distortion by comparing the Mel-Frequency Cepstral Coefficients (MFCCs) of the reconstructed and the ground truth speech, which ranges from 0 to infinity, with lower scores indicating higher speech quality.

\textbf{Multi-Resolution Mel-Spectral Loss (MEL)}: This metric evaluates the quality of reconstructed speech by measuring the distance between Mel-spectrograms at multiple resolutions. MEL scores range from 0 to infinity, with lower values indicating better speech quality.

\textbf{Word Error Rate / Character Error Rate (WER/CER)}:
These metrics assess the intelligibility of reconstructed speech by using an ASR model to transcribe the audio. The WER and CER reflect the percentage of incorrectly recognized words and characters, lower values indicate higher intelligibility.

\subsection{Overall Performance}\label{sec:overall_performance}

\subsubsection{Performance}
\Table{overall} presents the overall performance of \system across different test sets, which demonstrates efficient speech reconstruction performance across both seen and unseen speakers. On the seen-victim dataset ($\mathbb{D}1$), our model achieved an average FWSegSNR, STOI, MCD, and MEL of 9.43dB, 0.80, 5.18, and 2.09, respectively. Even on the unseen-user dataset $\mathbb{D}2$, the MCD was as low as 5.98. As pointed in \cite{yan2019feasibility}, an MCD below 8 indicates good perceptual speech quality. Therefore, we can conclude that \system delivers high-quality reconstructed speech. Although a slight performance drop is observed on the dataset $\mathbb{D}2$, the overall performance remains strong, showcasing \system's good generalization ability.

\begin{table}[h]
\centering
\caption{Overall Performance.}
% \vspace{-0.08in}
\setlength{\tabcolsep}{2.45pt}
  \begin{tabular}{c|cccc|ccccllll}
    \toprule
   \multirow{2}{*}{}  & \multicolumn{4}{c|}{Seen Victims} & \multicolumn{4}{c}{Unseen Victims}\\
& FWSegSNR$\uparrow$  & STOI$\uparrow$ & MCD$\downarrow$  & MEL$\downarrow$ & FWSegSNR$\uparrow$  & STOI$\uparrow$ & MCD$\downarrow$  & MEL$\downarrow$  \\
\midrule
Original Vibration Signal & 2.45 dB & 0.43 & 22.33   & 6.98  &    2.58 dB  &0.43  &21.54  &6.55   \\
\textbf{\system (Ours)}      & 9.43 dB    &0.80 &5.18 & 2.09   & 8.44 dB  &0.80 &5.93 &2.32 \\
\bottomrule
\end{tabular}
\label{tab:overall}
\vspace{-0.1in}
\end{table}

\begin{small}
\begin{table}[h]
 \newcommand{\tabincell}[2]{\begin{tabular}{@{}#1@{}}#2\end{tabular}}
\centering
  \caption{Comparison with prior work on vibration-based eavesdropping.}
     % \vspace{-0.1in}
  \label{tab:comp}
  \setlength{\tabcolsep}{2pt} %4pt
  \begin{tabular}{c|c|c|c|c|c|cllll}
    \toprule
                     & \tabincell{c}{Sensing\\ Target}  &  \tabincell{c}{Victim's\\ Data}   & \tabincell{c}{Sensing\\Content}  & \tabincell{c}{Required\\Inference Data} & \tabincell{c}{Through\\Wall}  & Performance\\
   \hline
   \tabincell{c}{mmSpy \cite{basak2022mmspy}\\ @S\&P'22}    &  Smartphone & Light  & Digits &mmWave & $\times $   &  Accuracy: 83\%  \\
   % \hline
   % Wavesdropper[IMWUT'22]\cite{wang2022wavesdropper}    & Speaker  &  Heavy &  \tabincell{c}{57\\ words} &  $\surd$&\\
    \hline
   \tabincell{c}{mmEve \cite{wang2022mmeve}\\ @MobiCom'22}    & Smartphone  & Light  &  \tabincell{c}{Harvard \\Sentences \cite{demonte2019harvard}}  &mmWave & $\times$  &   PSNR: 20 dB\\
   \hline
      \tabincell{c}{mmEavesdropper\cite{feng2023mmeavesdropper} \\@INFOCOM'23}    & Loudspeaker  & Heavy   &   \tabincell{c}{Digits \& \\Letters} &mmWave & $\surd$  &  PSNR: 15 dB \\
   \hline
      \tabincell{c}{VibSpeech \cite{wang2024vibspeech} \\ @Security'24}    &  \tabincell{c}{Loudspeaker$^{\ast}$\\Passive films} &  Light  & Speech & \tabincell{c}{mmWave\& \\speech} & $\surd$   & \tabincell{c}{w Voice: FWSegSNR: 5.4 dB\\ MCD: 3.9\\w/o Voice: FWSegSNR: -2.4 dB\\MCD: 9.7} \\
   \hline
      \tabincell{c}{\textbf{mmSpeech} \\\textbf{(Ours)}}  &  Passive films &  None  & \tabincell{c}{Unseen \& \\ unconstrained\\speech} &mmWave & $\surd$  &  \tabincell{c}{\textbf{FWSegSNR: 9.43 dB}\\ \textbf{MCD: 5.18}} \\
  \bottomrule
  \multicolumn{6}{l}{$\ast$ primary focus}
\end{tabular}
\vspace{-0.1in}
\end{table}
\end{small}

\subsubsection{Comparison}
To the best of our knowledge, no existing mmWave-based eavesdropping method operates under the same conditions as mmSpeech. Differences span several dimensions, including application scenarios, system capabilities, and radar configurations (\eg, sampling rate, chirp duration, and chirps per frame). We compare mmSpeech against representative baselines in \Table{comp} and highlight key distinctions.
Specifically, earlier methods such as mmSpy \cite{basak2022mmspy}, mmEve \cite{wang2022mmeve}, and mmEavesdropper \cite{feng2023mmeavesdropper} were limited to reconstructing fixed phrases and did not evaluate performance on unseen speakers, limiting their practicality in real-world applications. The SOTA method, VibSpeech \cite{wang2024vibspeech}, supports unconstrained sentence reconstruction but requires a small fragment of target user's speech data to achieve good results (FWSegSNR of 5.4 dB, MCD of 3.9). Without access to the target's voice data, its performance drops significantly (FWSegSNR of -2.4 dB, MCD of 9.7). In contrast, our method requires no prior knowledge of the victim's speech and can reconstruct unconstrained speech using only mmWave signals. We achieve the FWSegSNR of 9.43 dB and an MCD of 5.18—substantially outperforming VibSpeech under the same conditions and competitive to its performance when fused with audio data. In a nutshell, our approach significantly advances both performance and generalizability compared to existing vibration-based speech reconstruction methods.

\begin{figure*}[h]
\centering
\begin{minipage}[t]{\linewidth}
\centering
\subfigure[]
{\includegraphics[width=.38\textwidth]{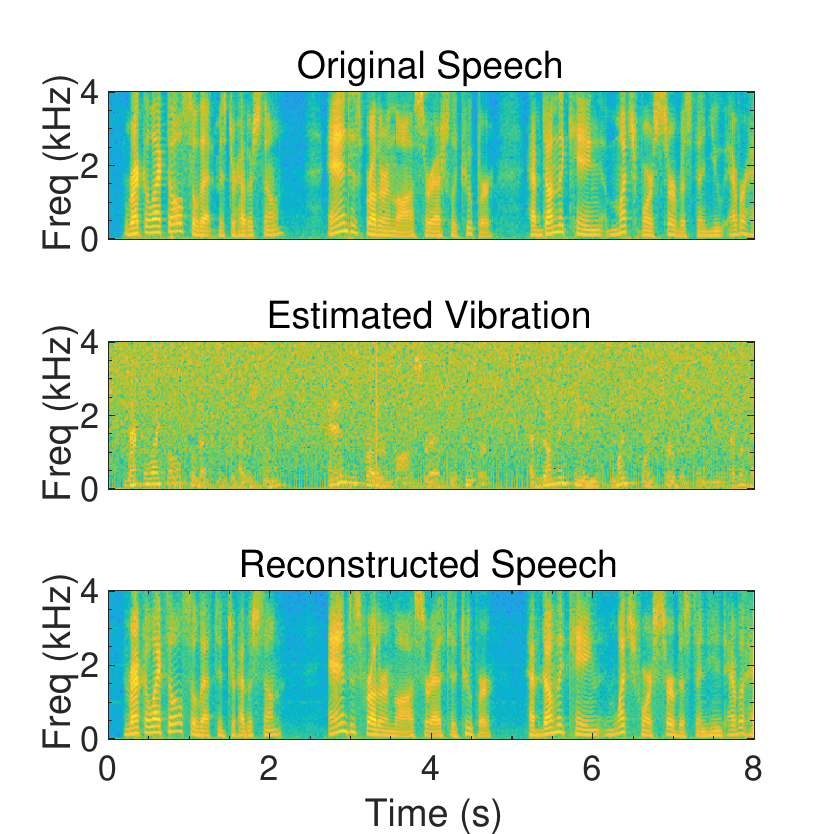}}
\subfigure[]
{\includegraphics[width=.38\textwidth]{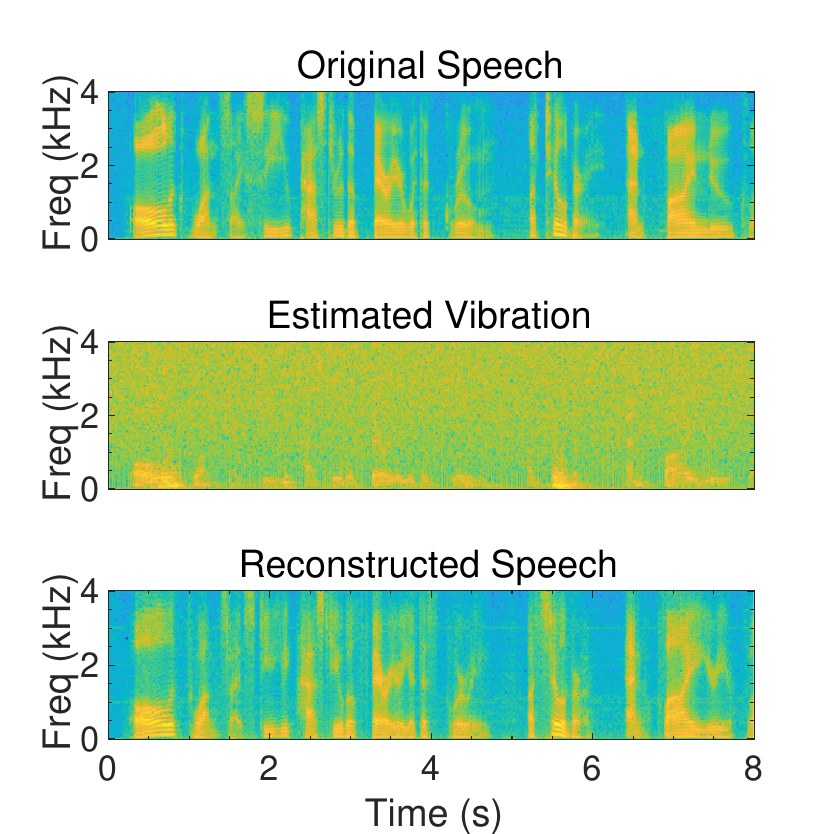}}
\vspace{-0.1in}
\caption{Visualization of the reconstructed audios. (a) `recollect also that mr heatherstone and his brother in law, sir ashley cooper, have done the king much more service than you ever'. (b) `all at once, i see you pass through the barrier with a groom, a tilbury and fine new clothes'. }
\label{fig:exp_example}
\end{minipage}
\vspace{-0.1in}
\end{figure*}

\subsubsection{Visualization}
We visualize two examples of spectrogram from the original speech, mmWave-estimated vibration, and our DNN reconstructed speech in \Figure{exp_example}. 
As shown in the top subfigure of \Figure{exp_example}(a), for speech segments with rich high-frequency content, the mmWave radar effectively captures vibrations above 1.5 kHz, as illustrated in the middle subfigure. Leveraging these high-frequency components in the vibration signals, the reconstructed speech successfully restores high-frequency details to a certain extent. On the other hand, for speech with blurred frequency components as the orgitnal speech seen in \Figure{exp_example}(b), the mmWave-captured vibrations also lack detail in the low-frequency range. Nevertheless, our DNN network is capable of compensating for these missing details through learning, as demonstrated by the well-recovered low-frequency elements in the reconstructed speech. These observations indicate that \system can effectively restore fine speech details from mmWave-derived vibration signals.

\subsubsection{Subjective Assessment}\label{sec:subjective_assessment}
To further evaluate the perceptual quality of our reconstructed speech, we conducted a subjective assessment using the Mean Opinion Score (MOS) test \cite{loizou2011speech}—one of the most widely adopted methods for speech quality evaluation. We invited 20 volunteers (aged 22 to 36) to rate the quality of speech samples on a five-point MOS scale, where 5 indicates `Excellent' and 1 denotes `Bad' quality.
We randomly selected 50 reconstructed speech samples from the testing set of $\mathbb{D}1$. Prior to the evaluation, each volunteer was asked to listen to six original speech samples (also from $\mathbb{D}1$) to establish a baseline for comparison. During the test, they rated the reconstructed samples without knowing they were generated using mmWave radar.

The results showed average scores ranging from 3.2 to 4.9 across the 50 samples. Among them, 26 were from male speakers and 24 from female speakers, with average ratings of 4.22 and 4.06, respectively. This slight difference aligns with the intuition that female speech, which contains more high-frequency components (>3 kHz), is more susceptible to degradation in mmWave-based sensing. Overall, \system achieved a strong average MOS of 4.14, confirming its ability to reconstruct intelligible speech suitable for eavesdropping across various conditions.
In addition, sample reconstructed by \system across various conditions—including seen speakers, unseen speakers, and through-wall scenarios—are available at \url{https://mmspeech.github.io/mmSpeech_demo/}. Note that all examples are unseen sentences drown only from our testing set.

\begin{table}[h]
 \newcommand{\tabincell}[2]{\begin{tabular}{@{}#1@{}}#2\end{tabular}}
\begin{minipage}{0.56\textwidth}
% \centering
  \caption{MOS rating scale.}
\setlength{\tabcolsep}{2.5pt}
 % \vspace{-0.1in}
  \begin{tabular}{cccl}
    \toprule
        Rating & Speech Quality  & Level of Distortion  \\
    \hline
    5    &   Excellent  & Imperceptible  \\
    4    &   Good  & Just perceptible, but not annoying \\
    3    &   Fair  & Perceptible and slightly annoying \\
    2    &   Poor & Annoying, but not objectionable  \\
    1    &   Bad  & Very annoying and objectionable  \\
  \bottomrule
\end{tabular}
\label{tab:exp_MOS}
\end{minipage}
\begin{minipage}{0.4\textwidth} % 调整宽度比例
% \centering
\caption{ASR performance.}
\setlength{\tabcolsep}{2pt}
 % \vspace{-0.1in}
  \begin{tabular}{cccl}
    \toprule
        & WER  & CER  \\
    \hline
    Real Speech    &   7.24  & 5.21  \\
    Initial mmWave Vibration &  86.79 & 65.21 \\
    Reconstructed Speech    &    69.92  & 47.71 \\
    \tabincell{c}{\textbf{\system} \\(with Fine-tuned ASR)}  &  34.20 & 18.95  \\
  \bottomrule
\end{tabular}
\label{tab:exp_ASR}
\end{minipage}%
% \hfill % 填充两个minipage之间的水平空间
\end{table}
% \end{center}

\subsubsection{Automatic Speech Recognition}
As demonstrated, our reconstructed speech achieves SOTA performance on objective metrics and is perceptually close to natural speech. Building on this, we aim to further test whether \system recovered speech enable automatic content understanding by applying standard Automatic Speech Recognition (ASR) systems. However, directly applying existing ASR models results in only moderate recognition accuracy. We analyze that this discrepancy arises from acoustic mismatches between reconstructed and real speech. Although these discrepancies do not alter semantic content, they introduce a distribution shift that impacts ASR models trained exclusively on real speech.

To address this issue, we adopt a transfer learning strategy. Specifically, we fine-tune the pre-trained SenseVoice \cite{gao2020san}, an ASR model trained on standard datasets with audio frequencies up to 8 kHz. SenseVoice uses an encoder-decoder architecture, where the encoder extracts acoustic features from the input audio, and the decoder generates textual transcriptions by leveraging both acoustic and contextual cues.

Since our reconstructed speech retains semantic consistency with real speech and differs only in acoustic representation, we design a selective fine-tuning approach: we keep the decoder frozen to preserve its language modeling capabilities, while we fine-tune the encoder to adapt to the characteristics of our reconstructed speech. This approach effectively bridges the domain gap and improves recognition accuracy.

We then evaluate the speech-to-text performance using the fine-tuned ASR model. To ensure fairness and maintain generalizability, only reconstructed speech from our training set of $\mathbb{D}1$ is used for fine-tuning. The model is then evaluated on the testing set with speech samples distinct from those used in training, to further verify the quality of our reconstructed speech. As shown in \Table{exp_ASR}, fine-tuning significantly boosts recognition performance: the CER drops from 47.71 to 18.95, and the WER decreases from 69.92 to 34.20. These results confirm the effectiveness of our approach and highlight the potential of \system for both eavesdropping and automatic content understanding.

\subsection{Micro-benchmarks}

\subsubsection{Vibrating Objects}

The quality of estimated vibrations is closely tied to the properties of the material used. \Section{sensing} has examined the frequency response characteristics of various materials and highlighted the differences in their frequency responses. To further investigate how vibrating material affects reconstructed speech quality, we evaluated and compared the performance using tinfoil and PET.

As shown in \Table{material}, PET outperforms tinfoil (most prior approaches used) in both the quality of the captured vibration signals and the reconstructed speech. Specifically, the vibration signal from PET achieved a FWSegSNR of 2.45 dB, which improved by 4.55 dB to 9.43 dB after applying our DNN model, while tinfoil achieved only 2.81dB. In terms of final reconstructed speech quality, the MCD for PET was 5.18 compared to 6.99 for tinfoil. These results indicate that PET is superior to tinfoil for speech eavesdropping using mmWave sensing. All subsequent experiments use PET as the default vibrating material for performance evaluation.

\begin{table}[t]
\centering
  \caption{Impact of vibrating materials.}
    % \vspace{-0.1in}
  \label{tab:material}
  \begin{tabular}{cccccllll}
    \toprule
&   FWSegSNR$\uparrow$  & STOI$\uparrow$ & MCD$\downarrow$  & MEL$\downarrow$  \\
   \hline
     PET  & 9.43dB (4.55$\uparrow$) &	0.80 (0.35$\uparrow$)	&5.18 (19.90$\downarrow$)	&2.09 (5.11$\downarrow$) \\
     Tinfoil              &  7.36dB	(6.98$\uparrow$) &0.77 (0.37$\uparrow$) &6.99	(17.15$\downarrow$) &	2.38 (4.89$\downarrow$)   \\
  \bottomrule
\end{tabular}
% \vspace{-0.15in}
\end{table}

\begin{figure*}[t]
\centering
\begin{minipage}[t]{\linewidth}
\centering
\subfigure[]
{\includegraphics[width=\textwidth]{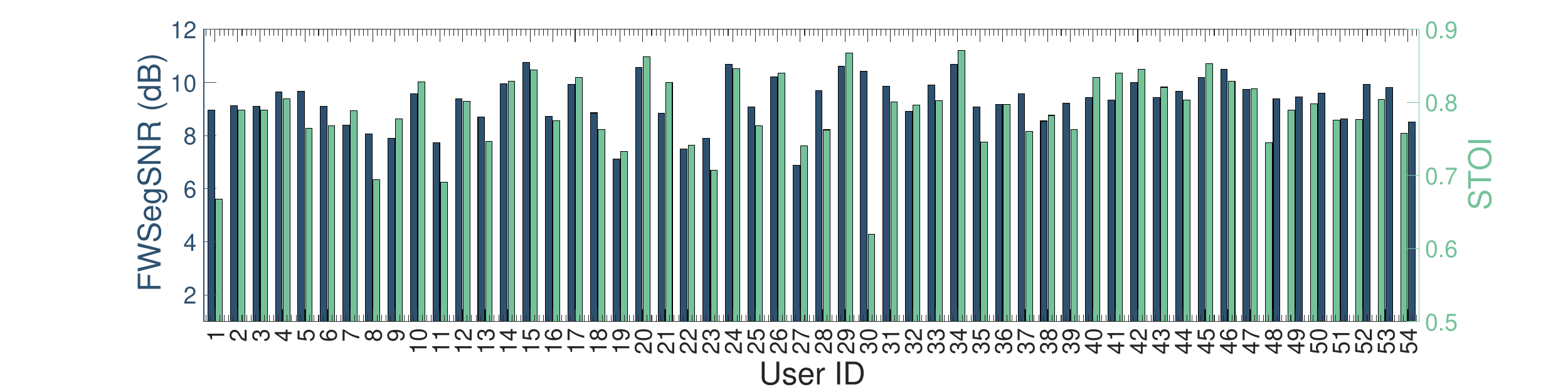}}
\subfigure[]
{\includegraphics[width=\textwidth]{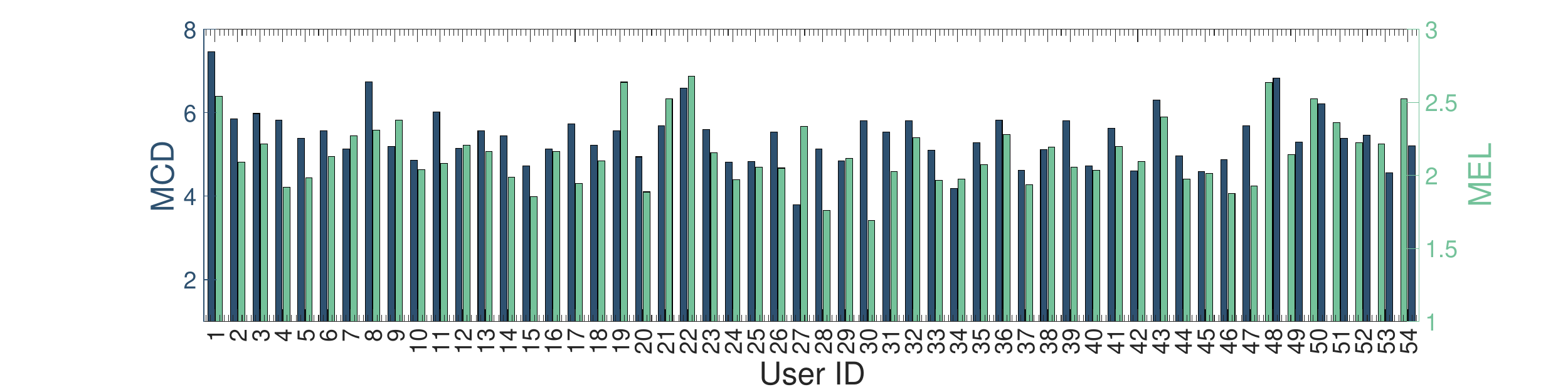}}
\caption{The performance of each individuals.}
\label{fig:exp_users}
\end{minipage}
% \vspace{-0.15in}
\end{figure*}

\subsubsection{Various Users}

Due to individual differences in vocal characteristics, the quality of vibration signals captured by the mmWave radar can vary across different speakers, which could affect the quality of reconstructed speech. For example, for individuals with higher-pitched voices, vocal energy tends to concentrate in higher frequency bands. However, displacement caused by high-frequency components that exceed the sensing range of the material may be attenuated. Additionally, mmWave signals tend to exhibit more noise in these high-frequency ranges, which may further degrade sensing performance. 

\Figure{exp_users} shows the speech reconstruction quality across users, where users \#0–\#46 represent seen users, and users 47–53 are unseen users. As illustrated, there is indeed notable variation in reconstruction performance across users. The gap in FWSegSNR between the best and worst results is around 3.55 dB, while the MCD difference reaches 3.8.
Nevertheless, even for the worst-performing user, the reconstructed speech achieves an FWSegSNR of 7.0 dB and MCD below 7.8, indicating that the reconstructed audio remains intelligible. Moreover, comparing the results between seen and unseen users shows that while the average performance for unseen users is slightly lower, there is no drastic degradation, demonstrating \system's strong generalization ability. These findings highlight the robustness of \system across different users.

\subsubsection{Sensing Distance and Angle}

\begin{figure*}[h]
\centering
\begin{minipage}[t]{.49\linewidth}
\centering
\subfigure[]
{\includegraphics[width=.49\textwidth]{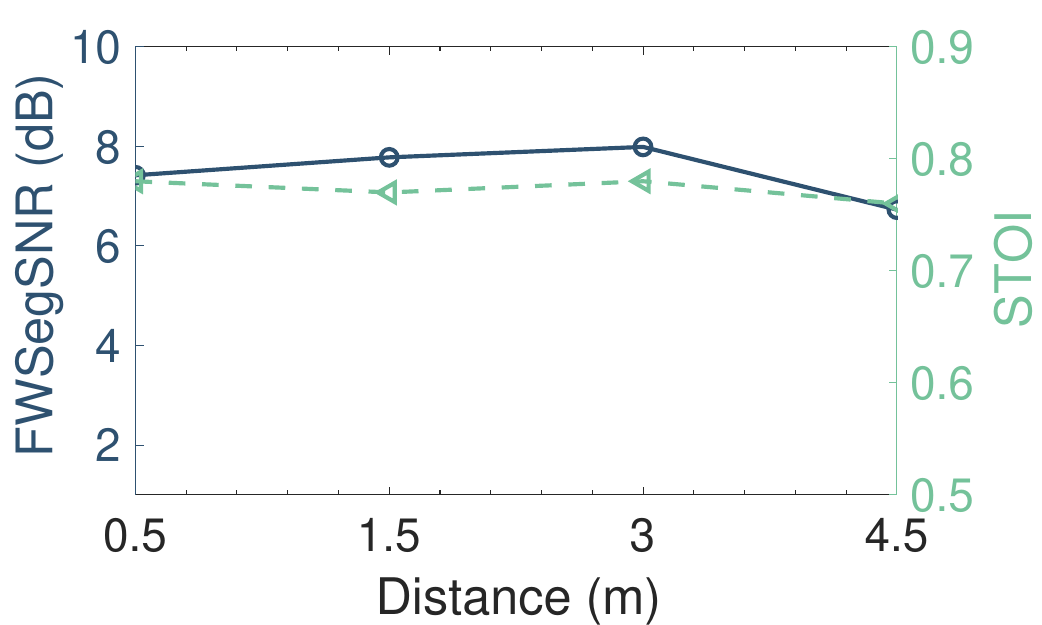}}
\subfigure[]
{\includegraphics[width=.49\textwidth]{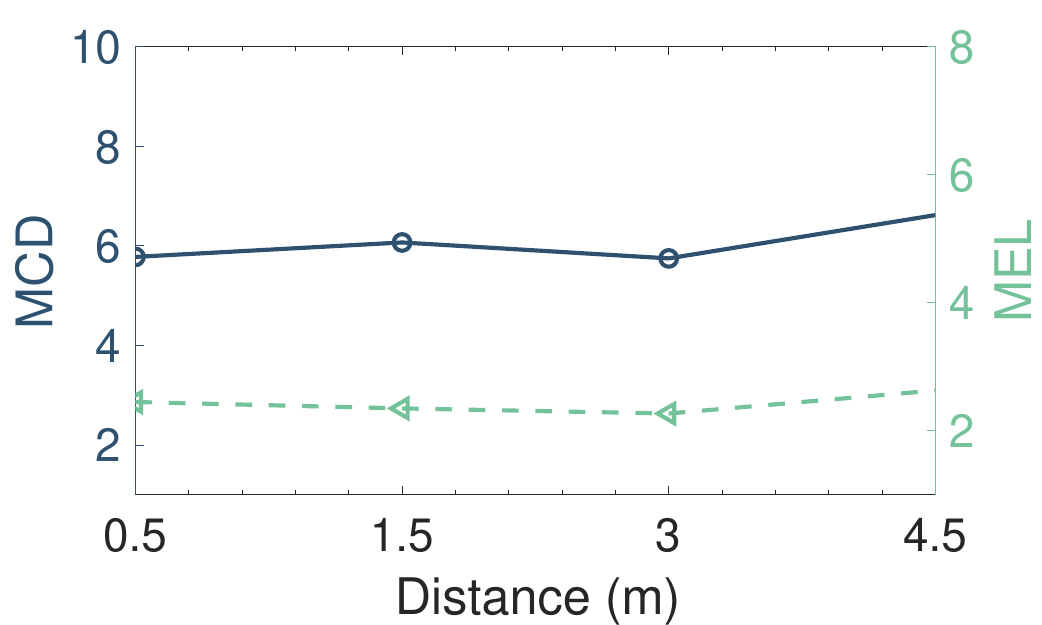}}
\caption{Impact of sensing distance.}
\label{fig:exp_distance}
\end{minipage}
\begin{minipage}[t]{.49\linewidth}
\centering
\subfigure[]
{\includegraphics[width=.49\textwidth]{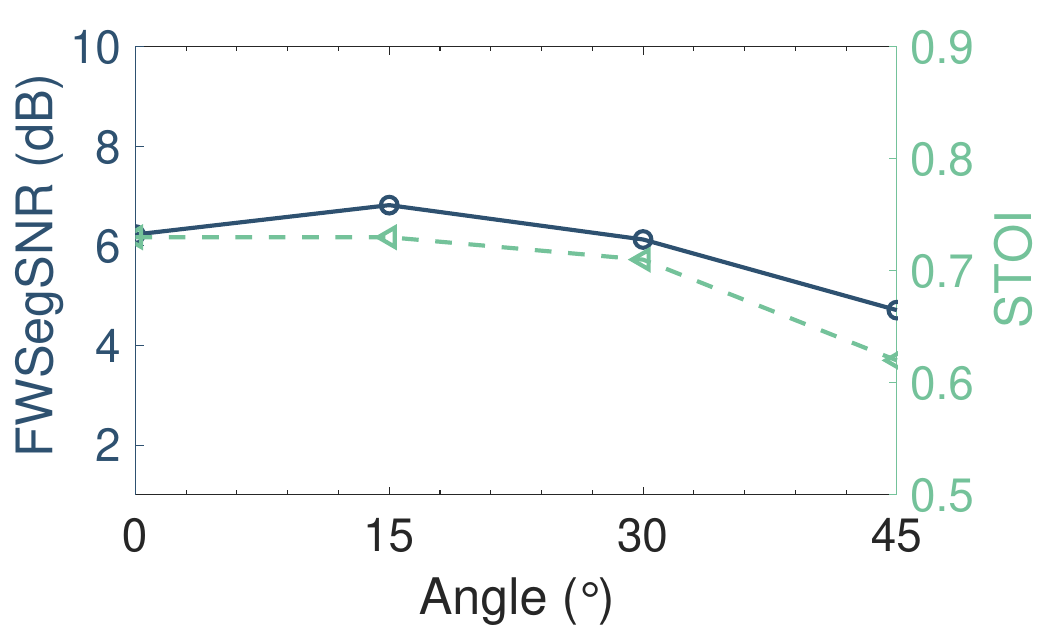}}
\subfigure[]
{\includegraphics[width=.49\textwidth]{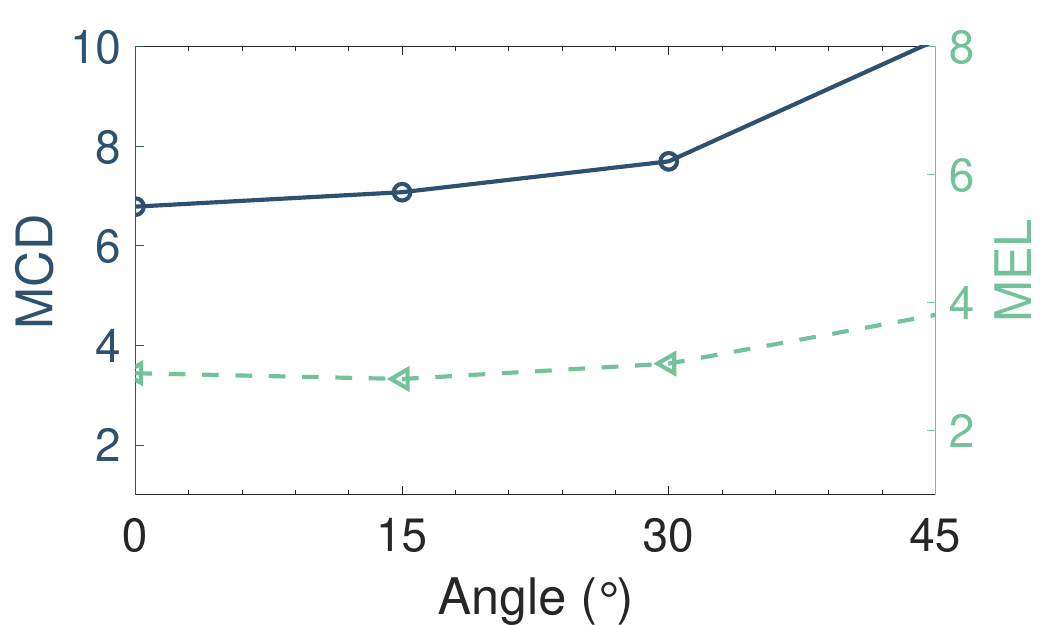}}
\caption{Impact of sensing angle.}
\label{fig:exp_angle}
\end{minipage}
% \vspace{-0.15in}
\end{figure*}

To investigate how the distance between the radar and the vibrating material affects speech reconstruction, we collected mmWave signals at various distances while keeping all other conditions constant. Specifically, measurements were taken at four distances: 0.5 m, 1.5 m, 3.0 m, and 4.5 m.
\Figure{exp_distance} presents the performance. The results show that within a 3-meter range, increasing the distance does not lead to a noticeable decline in speech reconstruction quality. However, when the distance reaches 4.5 m, a slight performance degradation is observed. Despite this, the reconstructed speech remains intelligible, with a FWSegSNR of 6.73 dB, MCD of 2.63, STOI of 6.62, and MEL of 2.63.
These findings demonstrate the robustness of our proposed system to variations in distance.

On the other hand, since mmWave signals tend to degrade significantly at larger angles, we further investigated the performance under varying angles. Specifically, we vary four angles, \ie, 0$^{\circ}$, 15$^{\circ}$, 30$^{\circ}$, and 45$^{\circ}$, while keeping all other conditions constant.
The results shown in \Figure{exp_angle} tell that as the angle increases from 0$^{\circ}$ to 30$^{\circ}$, there is no clear degradation in the quality of reconstructed speech. Although the FWSegSNR score worsens slightly, the MEL metric actually improves, suggesting a complex interplay between angle and signal quality. However, at 45$^{\circ}$, a significant drop in speech quality is observed—most notably, the MCD rises to 10.1. The results indicate the practical angular limit for \system's effectiveness.

\subsubsection{Blockage Materials}

The mmWave signals can penetrate various obstacles, enabling speech reconstruction even through physical barriers. To assess the impact of different obstacles, we evaluated three types of blockage materials, \ie, glass, acrylic plate, and cardboard (\Figure{scene}(d)(e)(f)). 
The results, summarized in \Table{blockage}, show clear differences in how each material affects the quality of the captured vibration signals. Cardboard had the least impact, followed by glass. While glass caused a noticeable degradation in speech quality, the reconstructed audio remained intelligible, demonstrating \system's ability to function through soundproof glass.
Notably, our reconstruction network significantly enhanced the quality of speech penetrated through glass. It improved the FWSegSNR by 4.85 dB, raising it to 5.99 dB. This result underscores the network's ability to recover lost signal information and affirms the robustness of our proposed method in real-world obstructed scenarios.

\begin{table}[h]
\centering
  \caption{Impact of blockage materials.}
    % \vspace{-0.1in}
  \label{tab:blockage}
  \begin{tabular}{ccccllll}
    \toprule
        & FWSegSNR$\uparrow$  & STOI$\uparrow$ & MCD$\downarrow$  & MEL$\downarrow$  \\
   \hline

        Glass     & 5.99 dB (4.85$\uparrow$) &   0.69 (7.17$\uparrow$) &7.62 (27.3$\downarrow$) &3.01 (4.91$\downarrow$)  \\
      Acrylic plate   & 4.90 dB (4.35$\uparrow$) &  0.64 (8.57$\uparrow$) &9.00 (29.39$\downarrow$) &3.49 (4.85$\downarrow$) \\
      Cardboard       & 7.68 dB (5.38$\uparrow$) & 0.76 (5.70$\uparrow$) & 6.16 (22.64$\downarrow$) &2.42 (4.62$\downarrow$)\\
  \bottomrule
\end{tabular}
% \vspace{-0.15in}
\end{table}

\subsubsection{Sound Volume}

\begin{figure*}[h]
\centering
\begin{minipage}[t]{.49\linewidth}
\centering
\subfigure[]
{\includegraphics[width=.49\textwidth]{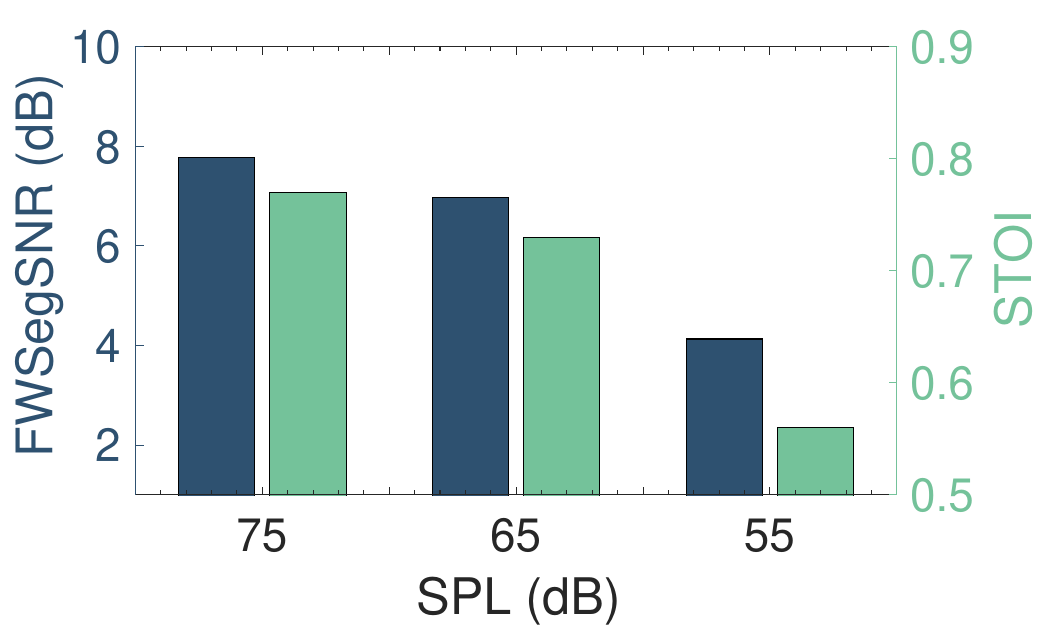}}
\subfigure[]
{\includegraphics[width=.49\textwidth]{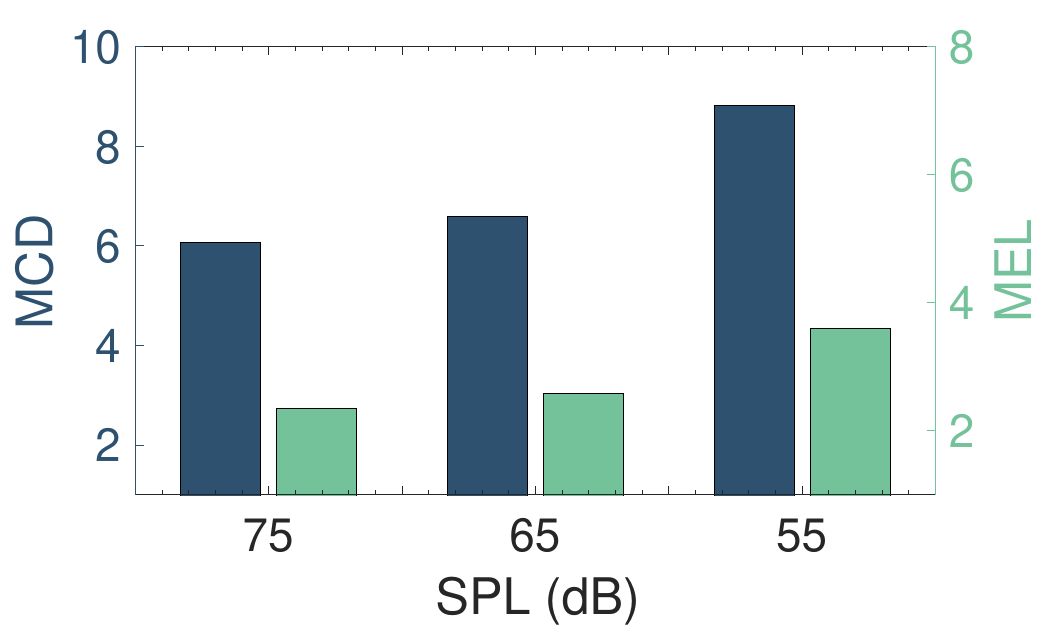}}
\caption{Impact of sound SPL.}
\label{fig:exp_SPL}
\end{minipage}
\begin{minipage}[t]{.49\linewidth}
\centering
\subfigure[]
{\includegraphics[width=.49\textwidth]{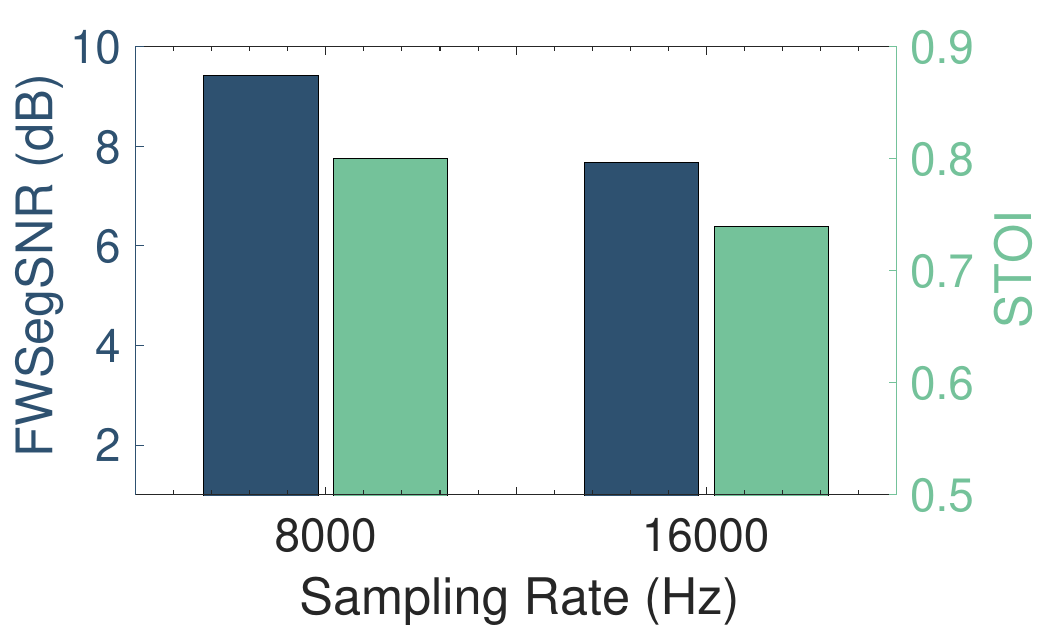}}
\subfigure[]
{\includegraphics[width=.49\textwidth]{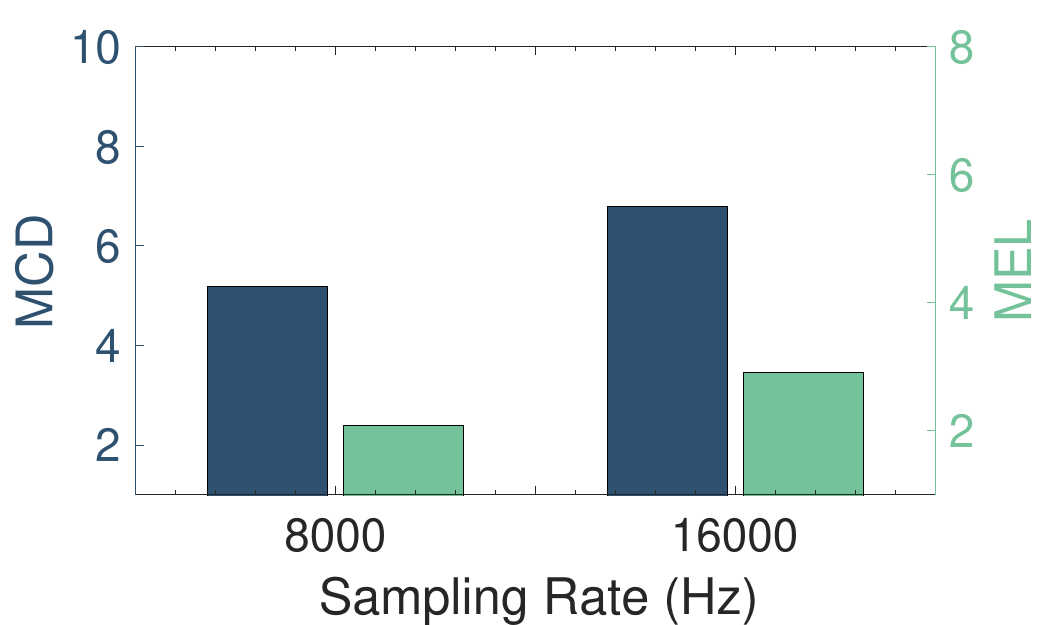}}
\caption{Impact of sampling rate.}
\label{fig:exp_samplingRate}
\end{minipage}
% \vspace{-0.15in}
\end{figure*}

Sound volume is a key factor influencing the amplitude of an object's forced vibration. To investigate how volume affects the reconstructed speech quality, we conducted experiments at three sound pressure levels (SPL): 75 dB, 65 dB, and 55 dB. For reference, the ambient noise level in the experimental environment was 45 dB SPL.
Intuitively, as volume decreases, the quality of the captured vibration signals is expected to degrade. The results presented in \Figure{exp_SPL} align with this expectation. When the volume dropped from 75 dB to 65 dB, the quality of speech showed no significant deterioration. However, when the volume further decreased to 55 dB, the degradation became much more obvious, \eg, MCD increased from 6.60 to 8.83.
These findings highlight that sound volume significantly impacts the quality of reconstructed speech: the louder the volume, the better the signal quality and the higher the reconstruction performance.

\subsubsection{Sampling Rate}

As discussed in \Section{vibration_sensing}, the sampling rate also plays a crucial role in determining the granularity of vibration estimation. To evaluate its impact on capturing speech-related vibrations below 4 kHz, we compare two sampling rates: 8000 Hz and 16000 Hz. As shown in \Figure{exp_samplingRate}, the 8000 Hz setting consistently outperforms 16000 Hz across all four metrics. This aligns with earlier observations in \Figure{de_samplingRate}, where higher sampling rates result in shorter chirps, reducing distance measurement resolution and thereby impairing both vibration sensing and speech reconstruction quality.

\subsection{Ablation Study}

To evaluate the effectiveness of each component in our proposed DNN model, we conducted ablation experiments on the $\mathbb{D}1$ dataset. The results, presented in \Table{exp_ablation}, illustrate the individual contributions of each module to the overall performance.
Eliminating the pretraining process resulted in a 0.3 dB drop in FWSegSNR, a 0.11 reduction in STOI, and increases of 0.11 in both MCD and MEL. These changes confirm that both synthetic data and the pretraining strategy significantly enhance model performance.
When the discriminator was removed, objective metrics showed only minor degradation. However, subjective listening revealed that the reconstructed speech sounded noticeably more mechanical and artificial. This highlights the discriminator's essential role in improving the perceptual quality of the output.
Removing the denoising module caused a substantial decline across all metrics: FWSegSNR dropped by 3.03 dB, MCD increased by 2.03, STOI fell by 0.13, and MEL rose by 0.53. These results underscore the denoising module's critical importance in maintaining speech reconstruction quality.

To further analyze component-level contributions, we examined the impact of removing individual loss functions.
$L_{mag}$: This loss minimizes the L1 distance between the generator's output spectrogram and the ground-truth. Its removal led to notable performance drops, particularly in FWSegSNR and MCD. Comparing the results of `-$L_{mag}$' with `- Denoising module', we find that $L_{mag}$ plays a key role in amplifying the denoising module's effectiveness.
$L_{Adv\_G}$: This reflects the case where MPD and MRD discriminators are not used for generator training. All metrics declined, and interestingly, performance was worse than in the case of removing the entire discriminator. This indicates that joint optimization of the generator and discriminator yields superior results.
$L_{FM}$: Removing the latent feature matching loss, which aligns intermediate features between real and generated samples, also degraded performance. This confirms its importance in guiding the model's internal representations.
$L_{MEL}$: Omitting this loss, which enforces similarity between multi-resolution mel-spectrograms of the reconstructed and reference speech, led to a noticeable degradation in reconstruction quality—particularly a 1.68 dB drop in FWSegSNR—highlighting its essential role in fine-grained spectral alignment.

\begin{table}[h]
\centering
  \caption{Ablation Study.}
  % \setlength{\tabcolsep}{1.35pt}
 % \vspace{-0.1in}
  \begin{tabular}{cccccl}
    \toprule
        & FWSegSNR$\uparrow$  & STOI$\uparrow$ & MCD$\downarrow$  & MEL$\downarrow$ \\
    \hline
    \textbf{\system}    &  9.43 dB  &0.80 &5.18 & 2.09   \\
    - Pretraining  & 9.13 dB & 0.79  &   5.29 & 2.20\\
    - Discriminator & 9.25 dB   & 0.79  & 5.82& 2.27\\
    - Denoising module &  6.40 dB   & 0.67  & 7.21& 2.62\\
    - $L_{mag}$   & 8.48 dB   & 0.76 & 5.81 & 2.27\\
    - $L_{FM}$    & 8.99 dB & 0.79  & 5.30   &2.13\\
    - $L_{Adv\_G}$ & 9.06 dB & 0.79  &5.33 &2.17 \\
    - $L_{MEL}$   &7.75 dB  & 0.77  & 5.86  & 2.41\\
  \bottomrule
\end{tabular}
\label{tab:exp_ablation}
% \vspace{-0.1in}
\end{table}

%% file: Body/discussion.tex
\section{Limitation and Discussion}\label{sec:discuss}

\subsection{Limitation} 
\system demonstrates the feasibility of reconstructing intelligible speech from mmWave-estimated vibrations. However, its performance is sensitive to the relative positioning between the radar and the vibration-sensitive surface (\eg, passive film). As detailed in Section 10.3.3, reliable speech recovery is achieved when the radar-to-surface angle is under 45$^{\circ}$ and the distance is within 4.5 m. Performance may degrade under larger angles or extended distances. Additionally, while mmSpeech supports through-wall sensing, its effectiveness depends on wall material and thickness; heavy or absorptive structures may hinder vibration capture.

\subsection{Defense Strategies} 
To mitigate the threat posed by malicious use of mmWave-based sensing, we suggest several countermeasures. (1) \textit{Vibration Damping}. Applying damping films or multilayer acoustic insulation to walls, windows, or nearby objects can suppress surface vibrations and reduce mmWave signal reflections, limiting the system's ability to capture intelligible signals. (2) \textit{Targeted Noise Injection.} Since mmSpeech primarily exploits vibrations in the 0–4 kHz band, introducing low-level or inaudible ambient noise in this range—while keeping it imperceptible to humans—will mask the vibration signature. (3) \textit{Speaker Placement.} Positioning loudspeakers away from reflective or vibration-prone surfaces, or placing them near materials with poor mmWave reflectivity, can significantly reduce information leakage through surface vibrations.

%% file: Body/cons.tex
\section{Conclusion} \label{sec:cons}

In this paper, we present \system, an end-to-end mmWave-based eavesdropping system that reconstructs intelligible speech solely from vibration signals induced by loudspeaker playback. Without requiring prior knowledge of the speaker and capable of functioning through walls, \system poses a practical threat to speech privacy. We addressed key challenges in fine-grained vibration sensing and speech reconstruction by optimizing material selection, radar sampling strategy, and designing a robust deep neural network. Extensive experiments using a commercial mmWave radar demonstrated that \system achieves SOTA performance and generalizes well across unseen speakers, unseen content, and various conditions, highlighting its real-world applicability and security implications.

%% file: Body/appendix.tex
\section{Details of the Loss Function}\label{sec:loss_appendix}

\subsubsection{MEL Loss}\label{sec:mel_loss}

The Mel spectrogram is a time-frequency representation of audio signals that aligns with human auditory perception. By applying a nonlinear frequency transformation known as the Mel scale, it converts raw audio into spectral features that better reflect how humans perceive sound. Incorporating Mel spectrogram-based constraints encourages our network to generate audio signals that are more perceptually natural.

To enhance frequency modeling across different time scales, we adopt a multi-resolution Mel-spectral loss (short for MEL loss $L_{MEL}$). $L_{MEL}$ not only improves training stability and signal fidelity but also accelerates model convergence \cite{kumar2023high}. Specifically, we compute the L1 loss between the generated and ground truth spectrograms across seven different Mel frequency configurations:
\begin{equation}\label{eq:L_denoise}
L_{MEL} = \sum_{j=1}^{7} \mathbb{E} \begin{Vmatrix} \varphi_j(s)-\varphi_j(G(x)) \end{Vmatrix}_1,
\end{equation}
where $\varphi(\cdot)$ denotes the Mel spectrogram computation function for the $j$-th group of setting. In our implementation, we use seven different sets of Mel spectrogram parameters to capture features at multiple resolutions. The number of Mel frequency bins is set to [5, 10, 20, 40, 80, 160, 320], while the corresponding window lengths are [32, 64, 128, 256, 512, 1024, 2048]. Each spectrogram is computed with a hop size equal to one-fourth of its window length, enabling the generator to capture both fine-grained and coarse spectral details across time.

% \blue{The Mel spectrogram is a time-frequency representation of audio signals that aligns with human auditory perception. By applying a nonlinear frequency transformation known as the Mel scale, it converts raw audio into spectral features that better reflect how humans perceive sound. Incorporating Mel spectrogram-based constraints encourages our network to generate audio signals that are more perceptually natural.}

% \blue{$L_{MEL}$ not only improves training stability and signal fidelity but also accelerates model convergence \cite{kumar2023high}. Specifically, we compute the L1 loss between the generated and ground truth spectrograms across seven different Mel frequency configurations. The number of Mel frequency bins is set to [5, 10, 20, 40, 80, 160, 320], while the corresponding window lengths are [32, 64, 128, 256, 512, 1024, 2048]. Each spectrogram is computed with a hop size equal to one-fourth of its window length, enabling the generator to capture both fine-grained and coarse spectral details across time.}

\subsubsection{Feature Matching Loss}\label{sec:feature_loss}

The feature matching loss is a similarity metric based on differences in feature representations extracted by the discriminator. It guides the generator by measuring the L1 distance between real and generated speech across intermediate layers of the discriminator, which is defined as:
\begin{equation}\label{eq:L_FM}
L_{FM} = \mathbb{E}_{(x,s)} \begin{bmatrix} \sum_{i=1}^{T} \frac{1}{N_i} \begin{Vmatrix} D_i(s)-D_i(G(x)) \end{Vmatrix}_1
\end{bmatrix},
\end{equation}
where $T$ is the number of layers in the discriminator, $N_i$ is the dimension of features at the $i$-th layer, and $D_i(s)$ and $D_i(G(x))$ refer to the feature representations of the real and generated speech, respectively, at layer $i$.
This loss function has been widely adopted in speech synthesis tasks \cite{lee2022bigvgan}\cite{kong2020hifi} for its effectiveness in improving training stability and enhancing generation quality.
Specifically, the discriminator extracts features from each intermediate layer. By aligning internal representations rather than just the final outputs, the generator is encouraged to produce more realistic and high-quality results.

% \blue{The feature matching loss is a similarity metric based on differences in feature representations extracted by the discriminator. As \Equation{L_FM} shown, it guides the generator by measuring the L1 distance between real and generated speech across intermediate layers of the discriminator.
% %
% This loss function has been widely adopted in speech synthesis tasks \cite{lee2022bigvgan}\cite{kong2020hifi} for its effectiveness in improving training stability and enhancing generation quality.
% %
% Specifically, the discriminator extracts features from each intermediate layer. By aligning internal representations rather than just the final outputs, the generator is encouraged to produce more realistic and high-quality results.}